%% file: main.tex
\documentclass[acmsmall,screen,nonacm]{acmart}
\usepackage{xcolor}
\usepackage{colortbl} 
\usepackage[ruled,linesnumbered]{algorithm2e}
\usepackage{graphicx}
\usepackage{ifthen} 
\usepackage{comment} 
\usepackage[only,llbracket,rrbracket]{stmaryrd}
\usepackage{mathtools} 
\usepackage{apxproof} 
\usepackage{listings}
\usepackage{sourcecodepro}
\usepackage{wrapfig}
\usepackage{booktabs}
\usepackage{centernot} 

\setcopyright{none}
\acmDOI{}
\acmISBN{}

\newboolean{anonymous}
\setboolean{anonymous}{true}

\newboolean{ourdraft}
\setboolean{ourdraft}{false}

\input{macros}

\usepackage[capitalise,noabbrev,nameinlink]{cleveref}
\crefname{algocf}{Algorithm}{Algorithms}
\crefname{definition}{Definition}{Definitions}
\crefname{rqcounter}{}{}
\creflabelformat{rqcounter}{#2RQ#1#3}

\begin{document}
\title{Soundness Checking of Taint Flow Models}

\author{Samarth Kishor}
\orcid{0009-0005-3795-3117}
\affiliation{%
  \institution{Amazon Web Services}
  \country{USA}
}
\email{samarkis@amazon.com}
\author{Victor Nicolet}
\orcid{0000-0002-3743-7498}
\affiliation{%
  \institution{Amazon Web Services}
  \country{USA}
}
\email{victornl@amazon.com}
\author{Joey Dodds}
\orcid{0009-0004-1534-6968}
\affiliation{%
  \institution{Amazon Web Services}
  \country{USA}
}
\email{jldodds@amazon.com}

\input{sections/0_abstract}

\maketitle

\input{sections/1_introduction}
\input{sections/2_example}
\input{sections/3_background}
\input{sections/4_approach}
\input{sections/5_instantiation}
\input{sections/6_evaluation}
\input{sections/7_related_work}
\input{sections/8_conclusions}

%
%
\newpage

\bibliographystyle{ACM-Reference-Format}
\bibliography{bibliography}

\newpage
\appendix
\input{appendix/llm}

\end{document}

%% file: macros.tex
\ifthenelse{\boolean{ourdraft}}{%
  \newcommand{\todo}[1]{{\color{red}\textbf{(TODO:} #1\textbf{)}}}
  \newcommand{\fix}[1]{\footnote{#1}}

  \newenvironment{newtext}{\color{blue}}{}

  \newcommand{\remove}[1]{\textcolor{red}{#1}}
  
  \newcommand{\victor}[1]{{\textcolor{magenta}{{\bf VN}: {#1}}}}
  \newcommand{\samarth}[1]{{\textcolor{purple}{{\bf SK}: {#1}}}}
}{%

  \newcommand{\todo}[1]{}
  \newcommand{\fix}[1]{}

  \excludecomment{removetext}
  \newcommand{\remove}[1]{}
  
  \newcommand{\victor}[1]{}
  \newcommand{\samarth}[1]{}
}

\newcommand{\defeq}{\stackrel{\mathrm{def}}{=}} 
\newcommand{\powerset}{\wp} 
\DeclarePairedDelimiter\denote\llbracket\rrbracket
\newcommand{\alloc}{\ensuremath{x \coloneq \textbf{new}}}
\newcommand{\assign}{\ensuremath{x \coloneq y}}
\newcommand{\load}{\ensuremath{x \coloneq \ast y.f}}
\newcommand{\store}{\ensuremath{\ast x.f \coloneq y}}
\newcommand{\callargs}{\ensuremath{{\downarrow}m(\overline{a})}}
\newcommand{\call}{\ensuremath{m(\overline{a})}}
\newcommand{\callretargs}{\ensuremath{\overline{x} \coloneq {\uparrow}\call{}}}
\newcommand{\callto}{\ensuremath{{\downarrow}\call{}}}
\newcommand{\callret}{\ensuremath{{\uparrow}\call{}}}
\newcommand{\incall}{\ensuremath{\text{In}_{\call{}}}}
\newcommand{\outcall}{\ensuremath{\text{Out}_{\call{}}}}
\newcommand{\inm}{\ensuremath{\text{In}_{m}}}
\newcommand{\outm}{\ensuremath{\text{Out}_{m}}}
\newcommand{\var}[1]{\ensuremath{{#1}}}
\newcommand{\accpath}[1]{\ensuremath{{#1}}}
\newcommand{\accpaths}[1]{\ensuremath{\pi_{#1}}}

\newcommand{\flowsto}[2]{\ensuremath{{#1} \rightsquigarrow {#2}}}
\newcommand{\trflowsto}[2]{\ensuremath{{#1} \rightsquigarrow^{\ast} {#2}}}
\newcommand{\nflowsto}[2]{\ensuremath{{#1} \mathrel{\centernot{\rightsquigarrow}} {#2}}}
\newcommand{\tto}{\ensuremath{\rightsquigarrow}}
\newcommand{\ntto}{\ensuremath{\mathrel{\centernot{\rightsquigarrow}}}}
\newcommand{\tmodel}[1]{\mcode|$\{$ #1 $\}$|}
\newcommand{\reach}[2]{\ensuremath{\mathit{reach}({#1},{#2})}}

\newcommand{\numRepos}{6}
\newcommand{\numTaintProperties}{16}
\newcommand{\numTopLevelModels}{97}
\newcommand{\numTopLevelFuncsAnalyzed}{2535}

\newcommand{\percentModelsSound}{93\%} 

\newcounter{rqcounter}

\newcommand{\rqitem}[2]{%
  \refstepcounter{rqcounter}%
  \item[{\bf RQ\therqcounter}] \label{#1}#2%
}

\newcommand{\code}[1]{\lstinline[style=gostyle,basicstyle=\footnotesize\ttfamily,mathescape=false]{#1}}
\newcommand{\mcode}{\lstinline[style=gostyle,basicstyle=\footnotesize\ttfamily,mathescape=true]}

\newcommand{\analysis}{\mathcal{A}}

\newcommand{\ptranalysis}{\mathcal{P}}

\lstdefinestyle{gostyle}{
  language=Go,
  basicstyle=\footnotesize\ttfamily,
  keywordstyle=\color{violet}\bfseries,
  commentstyle=\color{gray}\itshape,
  stringstyle=\color{teal},
  numberstyle=\tiny\color{gray},
  numbers=left,
  numbersep=8pt,
  tabsize=2,
  breaklines=true,
  breakatwhitespace=false,
  showstringspaces=false,
  frame=none,
  xleftmargin=2em,
  captionpos=b,
  aboveskip=1em,
  belowskip=1em,
  basewidth=0.5em,
  mathescape=true,
  morekeywords={nil},
  emph={len,errors,New,Write},
  emphstyle=\color{orange},
  emph={[2]ProcessReq,parseReq,logReq},
  emphstyle={[2]\color{green!50!black}}
}



%% file: sections/0_abstract.tex
\begin{abstract}
  Existing state-of-the-art static taint flow analyses for imperative programming languages can scale to large applications by using precise user-provided taint flow models of library methods. However, manually and precisely modeling a method's taint flows is tedious and potentially unsound. Furthermore, automatically modeling the method via an inter-procedural taint analysis can be inefficient. To solve this problem, we propose a guess-and-check approach: (1) an LLM agent that generates a precise taint flow model of a method and (2) a symbolic algorithm to check the soundness of the model. The algorithm deduces which taint flows \emph{must not} occur in the method for the LLM's taint flow model to be sound, and uses lightweight static analyses (e.g., type system and pointer analysis) to prove these must-not-flows. When these analyses are insufficient, the algorithm deduces maximally-general callee models and recursively verifies their soundness, avoiding a full inter-procedural taint analysis in most cases. Since a more precise model requires fewer must-not-flows to be verified, the precision of the LLM's model directly determines the efficiency of our approach. We evaluate our approach on
  \numTopLevelModels{}
  LLM-generated taint flow models for methods in \numRepos{} large Go codebases and prove the models sound for \percentModelsSound{} of the methods they cover.
  The proven-sound LLM-generated models are also precise, resulting in no new false-positives when proving taint flow properties.
\end{abstract}


%% file: sections/1_introduction.tex
\section{Introduction}
Reasoning about data flows in programs is useful for preventing security vulnerabilities such as logging secrets~\cite{CVETerraformLogSecrets24,CVEGitHubEnterpriseLogSecrets23} and for improving a developer's understanding of a program~\cite{weiserProgramSlicing1981a,sridharanThinSlicing2007a}. Static analyses such as taint analysis compute an over-approximation of these flows to prove security properties such as ``untrusted data from a call to method \code{getSecrets} (the source) never flows to an argument of \code{printf} (the sink)''. To \emph{prove} such properties, which we call taint flow properties, a static taint analysis must be \emph{sound} by over-approximating all the data flows from \code{getSecrets()} in the entire program.
A sound taint analysis is only useful when it is precise enough to prove all the taint flow properties with few-to-no false positives, i.e., taint flows that cannot occur in practice; and scalable, i.e., completes in a reasonable amount of time for large programs.


In order to be useful, a taint analysis must scale to such real-world applications. Even a simple ``hello world'' application in Java can transitively depend on 3,000 classes~\cite{tomanTamingStaticAnalysis2017}.
Precise whole-program taint analyses require hours and tens of gigabytes of memory to analyze large applications, even with sophisticated caching strategies~\cite{liScalingIFDSAlgorithm2021}.
To attempt to scale to large applications, compositional taint analyses use taint flow models of methods~\cite{banerjeeCompositionalTaintAnalysis2023}. A taint flow model of a method summarizes how tainted data flows from its inputs (e.g., parameters) to its outputs (e.g., return values) in \emph{all} calling contexts~\cite{arztStubDroidAutomaticInference2016}.
This enables the analysis to compute a method's model once and reuse it across different call sites.
Field-sensitive models summarize how taint flows from individual input object fields to output object fields, which can improve precision.
A model is sound if-and-only-if it over-approximates the taint flows in the method and its transitive callees for all calling contexts.


Previous work~\cite{arztStubDroidAutomaticInference2016,schubertLosslessPersistedSummarization2021}, which we call the constructive approach, use an inter-procedural compositional taint analysis to soundly and precisely model taint flows in methods.
Methods can manipulate objects, requiring the taint analysis to be field-sensitive by analyzing taint flows between individual object fields.
Some taint analyses are configurable to allow a user to sacrifice precision for performance; e.g., by disabling or limiting field-sensitivity~\cite{arztFlowDroidPreciseContext2014}. However, configuring an analysis is a manual operation and requires knowledge about the specific analysis and the application, which the user may not have. Furthermore, the resulting models may not be precise enough to avoid false-positives in real-world taint flow properties.

\paragraph{This work.}
Our insight is that instead of computing a sound and precise taint flow model via the constructive approach, we can instead use an external source such as a Large Language Model (LLM) to \emph{guess} a precise model which we then \emph{verify} is sound.
This addresses the efficiency and precision tradeoff: the method's taint flow model controls the level of precision that the analysis needs in order to verify soundness, not the method's implementation.

For example, consider an implementation of a method in Go for parsing and logging encoded data: \code{func parse(log *Logger, data []byte) error}. The type \code{[]byte} is a slice (heap-allocated mutable array) of bytes. The method \code{parse} and its transitive callees can contain thousands of statements that manipulate the input data in complex ways. However, its sound and maximally-precise taint flow model is simple: taint flows from the input data to the logger. A flow-, field-, and context-sensitive taint analysis for Go such as Argot~\cite{argot} can take minutes to compute this model, even though the model does not require field-sensitivity.
However, an LLM can guess this model in seconds, and proving that the model is sound does not even require a taint analysis; a field-insensitive pointer analysis is sufficient. If the memory pointed to by \code{log} is never written to the memory pointed to by \code{data} inside of \code{parse}, then we can conclude that data from \code{log} can never flow to \code{data}. Also, if the returned \code{error} value is a static constant, then data can never flow to it.

Our model generation approach equips an LLM with a set of analysis tools and tasks it with producing a candidate taint flow model for a method. Simply guessing the worst-case taint flow (all method inputs flow to all outputs) would likely lead to false-positives, so we equip the LLM with tools such as access to the method's documentation and source code to improve its precision.

LLMs can produce unsound candidate models.
To solve this problem, we use a novel soundness checking algorithm to \emph{verify} that the LLM's model over-approximates the possible taint flows within a method \( m \); i.e, for all possible inputs and outputs of \( m \) there is no flow from an input to an output in \( m \) or any of its transitive callees that is \emph{not} in the model.
Our modeling format has restrictions which we discuss in \Cref{sec:instantiation-model}.

The algorithm first computes the difference between the set of all possible flows between inputs and outputs of \( m \) and the model to deduce the taint flows that \emph{must not} occur in \( m \) for its model to be sound.
Then for each input-output flow that must not exist, called a ``must-not-flow'', it uses a series of \emph{efficient} static analyses to try and prove that the must-not-flow holds (i.e., that the output can never be tainted by the input) during the execution of \( m \).
Our static analyses use the language's type system and a pointer analysis.

When static analyses cannot directly prove that all the must-not-flows hold, and \( m \) calls other methods (callees), the algorithm computes the \emph{intra-}procedural taint flows in \( m \) using a standard taint analysis.
Akin to maximal specification synthesis~\cite{albarghouthiMaximalSpecificationSynthesis2016}, it then deduces which maximally-\emph{imprecise} models the callees of \( m \) need for the unproven must-not-flows and computed intra-procedural flows in \( m \) to hold.
The algorithm then recursively checks each deduced callee model.
Although it needs to perform a limited taint analysis in this case, we empirically show that the deduced callee models are imprecise enough that, in most cases, our static analyses can prove their soundness.
Therefore, in the general case, our soundness checking algorithm does not need to intra-procedurally analyze \emph{every} function that is transitively called from \( m \), which the constructive modeling approach would need to do.
The algorithm concludes that the LLM's model is sound iff there are no unproven must-not-flows in any callees.

This proof is relative to assumptions; in particular, the absence of unsafe memory operations and reflection. It also is relative to the application being analyzed, which motivates the need for efficiency, as all models will need to be re-checked sound each time the application changes. We discuss these assumptions in \Cref{sec:instantiation} and our approach's limitations in \Cref{sec:limitations}.

We instantiate our approach for the Go programming language and evaluate it across three~important dimensions of taint analysis: soundness, efficiency, and precision. We analyze \numRepos{} Go repositories with \numTaintProperties{} taint flow properties, automatically generating \numTopLevelModels{} models.
Applying our approach by generating the models, checking their soundness, and running the taint analysis, enabled us to prove 9 taint flow properties for which the baseline taint analysis (i.e., the analysis without models) timed out.
We show that our LLM-generated models lead to no additional false-positives compared to the baseline taint analysis.

\paragraph{Contributions.}
We make the following contributions:
\begin{itemize}
  \item We present an LLM agent to generate taint flow models.
  \item We present a novel approach to prove the soundness of a given taint flow model.
  \item We evaluate our approach on \numTopLevelModels{} taint flow models, empirically demonstrating our approach's efficiency.
\end{itemize}


%% file: sections/2_example.tex
\section{Running Example}

We consider a method \code{ProcessReq} implemented in Go which processes an HTTP request by parsing and asynchronously logging it. We provide the code for this method in \Cref{fig:ex-code}. \code{ProcessReq} takes as input a pointer to an HTTP request and a logger. It returns an \code{error} value (which is a pointer-like value in Go) if parsing the request fails. The returned \code{error} is \code{nil} (similar to \code{null}) if parsing and logging succeed. The callee \code{logReq} creates a new lightweight thread (called a ``goroutine'') via the \code{go} statement which eventually calls the \code{(*Logger).Write} method (a method called \code{Write} on the \code{*Logger} object) which logs its string parameter.

\begin{wrapfigure}{r}{.6\textwidth}
  \begin{lstlisting}
func ProcessReq(log *Logger, req *http.Request) error {
  body := parseReq(req)
  if len(body) == 0 {
    return errors.New("failed to parse req")
  }
  logReq(log, body, req)
  return nil
}

func parseReq(req *http.Request) string {
   // complex parsing implementation...
}
func logReq(log *Logger, body string, 
            req *http.Request) {
  go log.Write(req.Host + body)
}
  \end{lstlisting}
  \caption{Running example: an application with complex request parsing.}
    \label{fig:ex-code}
\end{wrapfigure}

This method uses features of Go which are difficult for a sound and precise taint analysis to analyze efficiently. The implementation of \code{parseReq} is complex and can take a long time to analyze. Furthermore, a flow-sensitive taint analysis cannot soundly summarize \code{logReq} without reasoning about concurrency. This makes \code{ProcessReq} a good candidate for our approach.

Instead of inefficiently \emph{constructing} a possibly-unsound but precise taint flow model for \code{ProcessReq} via a flow-sensitive inter-procedural taint analysis, we use an LLM agent to automatically generate a precise candidate model for the method, and then use our soundness checking algorithm to efficiently verify that the model is sound.

In theory, the user could configure the constructive approach to use a flow-insensitive taint analysis instead. However, that would require knowing ahead-of-time that the resulting model is precise enough. Our LLM agent generates a candidate taint flow model that is precise, yet our algorithm can use imprecise flow-insensitive analyses, which are sound in the presence of concurrency, to prove that the candidate model is sound.

We ask our LLM agent to generate a candidate taint flow model for \code{ProcessReq}. It responds with the model: data from the HTTP request \code{req} may taint the logger \code{log}, written as \tmodel{req $\tto$ log}.

The candidate model now needs to be proven sound; to do so, we run our soundness verification algorithm. The algorithm first deduces which taint flows must not hold in \code{ProcessReq} for the model to be sound. Then, it applies lightweight static analyses to prove that these \emph{must-not-flows} hold. If there are still some unproven must-not-flows, the algorithm deduces which models of the callees of \code{ProcessReq} (\code{parseReq}, \code{errors.New}, and \code{logReq}) satisfy the soundness of the LLM's generated model, and then recursively checks the soundness of each of these deduced callee models.

The most-general taint flow summary (we use ``model'' and ``summary'' interchangeably) states that taint from all method inputs flows to all method outputs. It must include \emph{all} possible inputs and outputs to a method in all calling contexts to over-approximate the possible summarized taint flows in a method. In Go, pointer-like parameters can be inputs as well as outputs to a method, so we include them in the outputs of the most-general summary. The difference between the most-general summary and the candidate model \tmodel{req $\tto$ log} is: \tmodel{req $\ntto$ ret, log $\ntto$ req, log $\ntto$ ret}, where \code{ret} is the return value. These are the taint flows that \emph{cannot} occur in \code{ProcessReq}, or any of its transitive callees, for the candidate model to be sound. If there are no must-not-flows, then the candidate model is the most-general summary, which is sound by construction.

Our algorithm is parameterized on a set of analyses that soundly prove that these must-not-flows hold. Each analysis takes as input set of must-not-flows and returns the must-not-flows it could not prove hold. Our instantiation of the algorithm for Go uses light-weight static analyses that analyze types and aliases. Crucially, these analyses are more efficient than the taint analysis of the constructive approach.

Our simplest analysis is the types analysis. In Go, non-pointer-like parameters are passed by-value on the stack. Even if a non-pointer-like parameter is modified inside of the method, the corresponding argument in the caller remains untouched; no data could possibly flow to it from any of the method's inputs. If a must-not-flow's output (e.g., \code{req} in \mcode|log $\ntto$ req|) is a non-pointer-like parameter, then the must-not-flow holds according to the types analysis; i.e., data \emph{never} flows from \code{log} to \code{req} in the method. The types analysis cannot prove any of these must-not-flows because the only output in the unproven must-not-flow set \tmodel{req $\ntto$ ret, log $\ntto$ req, log $\ntto$ ret} is \code{req}, which has a pointer type \code{*http.Request}. Therefore, the set of unproven must-not-flows is unchanged.

The next analysis is the immutability analysis, which uses a pointer analysis to prove that no data can ever flow to a pointer-like parameter output whose underlying memory is never modified (immutable) in the method. We require a pointer analysis instead of simply scanning for assignment statements because the taint analysis tracks taint flows through aliases as well as values. Due to imprecision in the pointer analysis, the immutability analysis incorrectly concludes that some statement in one of the transitive callees of \code{parseReq} modifies the underlying memory of \code{req}, so we cannot eliminate any must-not-flows to \code{req}.
Return values are a special case: they are always passed to the caller. However, if a return value is a constant value, then no data can flow to it.
The immutability analysis proves that \code{ret} is always a constant value: \code{nil} or an error string constant.\footnote{Our approach does not consider \emph{implicit} taint flows, so even though the return value depends on the condition \code{len(body) == 0}, our soundness verification algorithm concludes that no tainted data can flow to the return value.}
Thus, the set of unproven must-not-flows is now \tmodel{log $\ntto$ req}.

Next, the read analysis checks that no data can ever flow to an input that is never read (unread) in the method. Since the \code{logReq} method accesses \code{req} to call a method on it, the read analysis cannot prove the remaining must-not-flow.

\begin{figure}
  \includegraphics{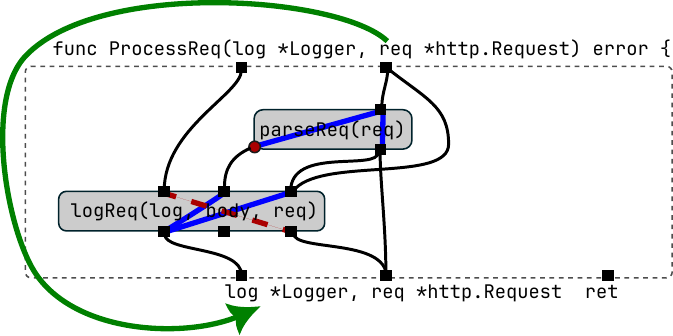}
    \caption{Taint flow graph representation of \code{ProcessReq}. The long arrow is the flow in the LLM's candidate taint flow model. Small square boxes below parameters or above arguments of call statements (inside the shaded rounded boxes) denote method inputs. Black boxes below arguments and dots below call statements (return values) denote method outputs. Black lines outside of the call statements are the computed intra-procedural taint flows. Solid lines inside of the call statements are the unknown input-output flows for each respective deduced callee model. The dotted line is a callee model flow which would violate the soundness of the candidate model.}
    \label{fig:ex-flows}
\end{figure}

We have no other static analyses to prove that the rest of the unproven must-not-flows hold. Thus, we must compute all the possible intra-procedural flows from each input of \code{ProcessReq} to its outputs, and deduce which callee taint flows satisfy these intra-procedural flows \emph{and} the unproven must-not-flows, for the candidate model to be sound.
The algorithm taints every input to \code{ProcessReq} (i.e., \code{log} and \code{req}) and runs the \emph{intra}-procedural taint analysis on \emph{only} \code{ProcessReq}, not any of its callees. This is in contrast to the constructive approach which must analyze taint flows in every callee.

\Cref{fig:ex-flows} represents the intra- and inter-procedural taint flows in \code{ProcessReq} as a graph. The taint analysis computes the intra-procedural flows (black lines in \Cref{fig:ex-flows}) soundly and efficiently because \code{ProcessReq} does not perform much computation itself. The intra-procedural analysis does not analyze any callees---it leaves ``call nodes'' representing calls to callees (gray boxes in \Cref{fig:ex-flows}). It proceeds by assuming every output to the call is tainted, which is necessary to ensure soundness.

We now know the intra-procedural taint flows within \code{ProcessReq}. However, the taint flow graph is still incomplete because we are missing edges that connect the inputs and outputs of each call node. We want to avoid performing a potentially expensive taint flow analysis on each callee, so instead our goal is to \emph{deduce possible models}. Furthermore, we want to \emph{maximize} the number of flows in the deduced callee models which in-turn minimizes the number of must-not-flows that the static analyses will need to prove in each callee. These deduced callee model flows are the blue lines inside of the call statements in \Cref{fig:ex-flows}. Our approach does not deduce a model for \code{errors.New} because it is unreachable from the inputs to \code{ProcessReq}.

The deduced callee models must not violate the soundness of the LLM's candidate model. For example, if \code{logReq} actually taints \code{req} with \code{log} (dotted red line in \Cref{fig:ex-flows}), perhaps by storing some previously-logged data inside of the request's byte buffer, then there would be a flow from the \code{logReq} call's input argument \code{log} to the call's output argument \code{req} to the output value \code{req} of \code{ProcessReq}. This extra flow from call output \code{req} to \code{ProcessReq} output \code{req} would make the LLM's candidate model unsound.
Therefore, we add the constraint that the deduced callee models must satisfy the unproven must-not-flows \emph{and} the known intra-procedural flows that we just computed.

Our algorithm deduces the following maximally-imprecise callee models:
\begin{enumerate}
  \item \code{parseReq}: \tmodel{req $\tto$ ret}
  \item \code{logReq}: \tmodel{body $\tto$ log, req $\tto$ log, log $\tto$ body}
\end{enumerate}

The deduced summary for \code{parseReq} is the most-general summary \tmodel{req $\tto$ ret} which is sound by construction. This is convenient because while the implementation of \code{parseReq} may be too complex for a precise taint analysis to summarize efficiently, simply using the sound most-general summary lets us avoid performing an expensive taint flow computation \emph{without losing any precision}.

The algorithm moves on to proving that the deduced model for \code{logReq} is sound. The most-general summary for \code{logReq} is \tmodel{log $\tto$ body, log $\tto$ req, body $\tto$ log, body $\tto$ req, req $\tto$ log, req $\tto$ body}. This gives us \tmodel{log $\ntto$ req, body $\ntto$ req, req $\ntto$ body} as the set of unproven must-not-flows. The types analysis proves that \mcode|req $\ntto$ body| holds because body is a non-pointer-like parameter (string). This leaves us with the set \tmodel{log $\ntto$ req, body $\ntto$ req}. The algorithm then performs the immutability analysis, which proves that the memory which \code{req} refers to is never modified by \code{logReq} or any of its transitive callees. This is sound even though \code{logReq} spawns a goroutine if the immutability analysis uses a flow-insensitive pointer analysis, which does not consider the order that statements are executed. Our approach assumes that the programs we analyze are free from data races. Under this data-race-freedom assumption, we conclude that \code{logReq} does not modify \code{req}, leaving us with no more unproven must-not-flows. The algorithm finally concludes that the LLM-inferred model of \code{ProcessReq}, \tmodel{req $\tto$ log}, is sound, without performing an inefficient taint analysis on any of its callees.


%% file: sections/3_background.tex
\section{Taint Semantics}

In this section, we define the semantics of taint flows for a simple imperative programming language.
The language we use is similar to the one in \textsc{Andromeda}~\cite{trippAndromedaAccurateScalable2013} and \textsc{FlowDroid}~\cite{fritzHighlyPreciseTaint2013}. It supports manipulating heap-allocated objects and calling methods.
We instrument the semantics of our language to track taint flows.
Finally, since our approach deals with the soundness of taint flow models, we use the instrumented semantics to formalize the definition of a sound taint flow model.

\subsection{Programming Language}\label{sec:simple-language}

\begin{figure}
  \centering
  \begin{minipage}{.49\textwidth}
    \begin{tabular}{ll}
      Name & Description \\
      \hline
      \( \mathbb{V} \) & Variables \\
      \( \mathbb{F} \) & Field identifiers \\
      \( \mathbb{S} \) & Statements \\
      \( \mathbb{M} \) & Methods \\
      \( \mathbb{L} \) & Memory locations \\
    \end{tabular}
  \end{minipage}
  \begin{minipage}{.49\textwidth}
    \begin{tabular}{ll}
      Name & Description \\
      \hline
      \( \mathbb{O} \defeq \mathbb{L} \cup \{\text{null}\} \) & Object values \\
      \( \mathbb{E} \defeq \mathbb{V} \to \mathbb{O} \) & Environment \\
      \( \mathbb{H} \defeq \mathbb{L} \times \mathbb{F} \to \mathbb{O} \) & Heap \\
      \( \Sigma \defeq \mathbb{E} \times \mathbb{H} \) & Program states
    \end{tabular}
  \end{minipage}
  \caption{Our programming language's domain.}\label{fig:domain}
\end{figure}

\Cref{fig:domain} shows the programming language's domain.
A program state \( \sigma \defeq \langle E, H \rangle \in \Sigma \) maintains the state of the environment and heap. The environment \( E \in \mathbb{E} \) maps program variables in \( \mathbb{V} \) to null-able object values in \( \mathbb{O} \).
The heap \( H \in \mathbb{H} \) maps memory locations in \( \mathbb{L} \) with field identifiers in \( \mathbb{F} \) to object values.
\begin{wrapfigure}{r}{.6\textwidth}
  \begin{tabular}{rlr}
    \multicolumn{2}{c}{Meaning} & Description \\
    \hline
    \( \denote{\alloc{}}\sigma \) & \(= \sigma[x \mapsto \ell \in \mathbb{L}.\ \ell \text{ is fresh}]\) & allocation \\
    \( \denote{\assign{}}\sigma \) & \(= \sigma[E(x) \mapsto E(y)]\) & assignment \\
    \( \denote{\load{}}\sigma \) & \(= \sigma[E(x) \mapsto H(\langle E(y), f \rangle)]\) & load \\
    \( \denote{\store{}}\sigma \) & \(= \sigma[H(\langle E(x), f \rangle) \mapsto E(y)]\) & store
  \end{tabular}
  \caption{Semantics for basic statements.}\label{fig:basic-semantics}
\end{wrapfigure}
The evaluation function \( \denote{\cdot} : \mathbb{S} \times \Sigma \to \Sigma \) evaluates a statement \( s \in \mathbb{S} \) from a state \( \sigma \in \Sigma \), producing state \( \sigma' \).
We present the language's basic statements in \( \mathbb{S} \) with their corresponding concrete semantics in \Cref{fig:basic-semantics}.
We use \( \powerset \) to denote the power-set.

\subsubsection{Method calls}
\label{sec:method-calls}

Our language supports method calls. The statement \( \callargs{} \) calls a callee method \( m \) with arguments \( a_0, a_1, \ldots, a_n \in \overline{a} \). The callee returns to the caller via the statement \( \callretargs{} \) with the return values \( x_0, x_1, \ldots, x_n \in \overline{x} \). For brevity, we omit \( \overline{x} \) when it is implied from the context.
We write \( \call{} \) to refer to the method call execution.

The call statement \( \callargs{} \) establishes the initial callee state by creating a fresh environment containing only the parameters bound to their corresponding argument values:
\[ \denote{\callargs{}}\sigma = \langle \langle p_1 \mapsto E(a_1), \ldots, p_n \mapsto E(a_n) \rangle, H \rangle \]
where \( p_i \) is the parameter corresponding to argument \( a_i \in \overline{a} \), and \( E \) is the environment from \( \sigma \).
The callee shares the heap \( H \) with the caller. Thus, any modifications to caller-accessible heap-allocated variables in the \emph{callee} are observable in the \emph{caller} after the callee returns.

\subsubsection{Program traces}

A \emph{program trace} \( \tau \defeq [ s_0, s_1, \ldots, s_n ] \) is an ordered sequence of program statements, where \( s_i \in \mathbb{S} \). The set of all possible traces of a given program is \( \mathrm{T} \). 
There exists a prefix relation \( \preceq \) on traces such that for traces \( \tau_0, \tau_1 \in \mathrm{T} \), we have \( \tau_0 \preceq \tau_1 \) iff \( \tau_0 \) is a prefix of \( \tau_1 \). The prefix relation allows us to reason about subsequences of program traces.
Given an initial state \( \sigma \), we inductively define the evaluation of a trace \( \text{eval} : \mathrm{T} \times \Sigma \to \Sigma^* \) as:
\[
  \text{eval}(\tau, \sigma) \defeq
  \begin{cases}
    [\sigma] & \text{if } \tau = [] \\
    [\sigma] \cdot \text{eval}([ s_1, \ldots, s_n ], \denote{s_0}\sigma) & \text{if } \tau = [ s_0, s_1, \ldots, s_n ]
  \end{cases}
\]

\subsection{Taint Flows}
\label{sec:taint-semantics}

We instrument the concrete semantics to track taint flows between \emph{access paths}. An access path, borrowed from \textsc{Andromeda}~\cite{trippAndromedaAccurateScalable2013}, is a symbolic representation of a heap location.
The set of all access paths is \( \Pi \defeq \mathbb{V} \times \mathbb{F}^* \).
We write an access path \( x.f_0.f_1._{\ldots}.f_k \) where \( \var{x} \in \mathbb{V} \) is a local variable and \( f_0, f_1, \ldots, f_{k} \) is a finite sequence of field identifiers where \( f_i \in \mathbb{F} \). Access paths are bounded to a predetermined length \( k \), which lets us soundly analyze recursive objects such as linked lists.
We write \( \accpath{x.f} \) for the access path referring to field \( f \) of variable \( \var{x} \).

The taint flow domain \( \mathbb{T} \defeq \powerset (\Pi \times \Pi) \) is a set of taint flows between access paths.
An \emph{instrumented concrete state} is a triple \(\sigma \defeq \langle E, H, T \rangle\), where \( T \in \mathbb{T} \) tracks taint flows.
We write \( \flowsto{\accpath{a}}{\accpath{b}} \) if \( \langle \accpath{a}, \accpath{b} \rangle \in \sigma_T \) when data from access path \( \accpath{a} \) flows to \( \accpath{b} \) in state \( \sigma \).
The flow \( \trflowsto{\accpath{a}}{\accpath{b}} \) means data from access path \( \accpath{a} \) \emph{transitively} flows to \( \accpath{b} \).


We extend the concrete semantics \( \denote{\cdot} : \mathbb{S} \times \Sigma \to \Sigma \) to propagate taint flows. \Cref{fig:taint-semantics} shows the instrumented semantics for basic statements, where \( \flowsto{\cdot}{\pi} \) means \( \exists \pi'. \langle \pi', \pi \rangle \in T \).
To resolve aliasing between access paths, we require a sound \emph{pointer analysis} \( \ptranalysis \subseteq \powerset(\Pi) \). Since our approach does not rely on a specific pointer analysis, we do not describe the implementation of \( \ptranalysis \). Let \( \text{pts}(\accpath{\pi}) \subseteq \powerset(\Pi) \) denote the points-to set of access path \( \accpath{\pi} \), representing all abstract locations that \( \accpath{\pi} \) may point to. Let \( \ptranalysis(\accpath{\pi}) \defeq \accpath{\pi} \cup \{\accpath{\pi'} \mid \text{pts}(\accpath{\pi}) \cap \text{pts}(\accpath{\pi'}) \neq \emptyset\} \) include all access paths that may alias \( \accpath{\pi} \), including itself.
The store statement \( \store{} \) uses the pointer analysis to taint all access paths that may alias \( \accpath{x.f} \) (including \( \accpath{x.f} \) itself) with \( \accpath{y.\varepsilon} \) (we use \( \varepsilon \) to signify the empty field).

\begin{figure}
  \label{fig:taint-semantics}
  \begin{tabular}{rlr}
    \multicolumn{2}{c}{Taint update} & Description \\
    \hline
    \( \denote{\alloc{}}\sigma \) & \(=_T \sigma_T\) & allocation \\
    \( \denote{\assign{}}\sigma \) & \(=_T \sigma_T \cup \{ \flowsto{\accpath{y.\varepsilon}}{\accpath{x.\varepsilon}} \mid \flowsto{\cdot}{\accpath{y.\varepsilon}} \in \sigma_T \}\) & assignment \\
    \( \denote{\load{}}\sigma \) & \(=_T \sigma_T \cup \{ \flowsto{\accpath{y.f}}{\accpath{x.\varepsilon}} \mid \flowsto{\cdot}{\accpath{y.f}} \in \sigma_T \}\) & load \\
    \( \denote{\store{}}\sigma \) & \(=_T \sigma_T \bigcup_{\accpath{a} \in \ptranalysis(\accpath{x.f})} \{ \flowsto{\accpath{y.\varepsilon}}{\accpath{a}} \mid \flowsto{\cdot}{\accpath{y.\varepsilon}} \in \sigma_T \}\) & store
  \end{tabular}
  \caption{Instrumented taint semantics for basic statements}
\end{figure}

For method calls, taint flows from parameters to arguments. Given a method call statement \( \callargs{} \) and state \( \sigma \):
\[
  \denote{\callargs{}}\sigma_T = \{ \flowsto{\accpath{a_i}}{\accpath{p_i}} \mid \flowsto{\cdot}{\accpath{a_i}} \in \sigma_T \}
\]
The arguments are the input set to the call: \( \text{In}_{\call{}} \defeq \{ \var{a_0}, \ldots, \var{a_n} \} \).

When the callee returns via \( \callretargs{} \) from final callee state \( \sigma \), taint flows back to the caller state \( \sigma' \):
\[
  \denote{\callretargs{}}\sigma_T = \sigma'_T \cup
  \begin{cases}
    \{ \flowsto{\accpath{r_i}}{\accpath{x_i}} \mid \flowsto{\cdot}{\accpath{r_i}} \in \sigma_T \} \\
    \{ \flowsto{\accpath{p_i}}{\accpath{a_i}} \mid \flowsto{\cdot}{\accpath{p_i}} \in \sigma_T \}
  \end{cases}
\]
where \( r_i \) are the return values from the callee and \( x_i \in \overline{x} \) are the variables receiving them in the caller.
Since all objects in our language are allocated on the heap, taint also flows from the parameters in the callee back to the arguments in the caller.
These return values are the output set from the call: \( \text{Out}_{\call{}} \defeq \{ \var{x_0}, \ldots, \var{x_n}, \var{a_0}, \ldots, \var{a_n} \} \).

\subsection{Taint Flow Model}

Our approach verifies the soundness of a given taint flow model. We first define a taint flow model in terms of our language and then define the soundness condition for a model.
Our taint flow model representation is inspired by \textsc{StubDroid}~\cite{arztStubDroidAutomaticInference2016} where the model \( \Phi_m : \Pi \times \Pi \) of a method \( m \) is the set of access path tuples representing \emph{caller-visible} flows from call inputs in \( \incall{} \) to outputs in \( \outcall{} \) for \emph{all possible} method calls.
We use ``model'' and ``summary'' interchangeably; usually, a model is given and checked against the implementation, whereas a summary is constructed from the implementation.

A program trace may contain multiple distinct calls to a method \( m \), each of which may add new taint flows to the caller's state. We first define a \emph{call segment prefix} of the trace, which is a trace prefix ending with the statements corresponding to the execution of a call to \( m \). We then define a taint summary for a single method call in terms of evaluating the call segment prefix, and finally generalize the definition for all possible method calls in the program.

\begin{definition}
  \label[definition]{def:call-segment-prefix}
  Given a program trace \( \tau \) and initial state \( \sigma_0 \), we define the following:
  \begin{itemize}
    \item The \emph{call segment prefix} is the prefix of trace \( \tau \) ending with the return statement \( \callret{} \) corresponding to the call statement \( \callto{} \):
      \[ \tau_{\callto{}} \defeq \tau' [\callto{}] {S_{\callto{}}} [\callret{}] \preceq \tau \]
    \item The \emph{call segment} \( S_{\callto{}} \) includes all the statements in \( m \) (and its transitive callees) in \( \tau \) between the call statement \( \callto{} \) and the return statement \( \callret{} \).
    \item The \emph{call state} is the state in the caller before evaluating the statement \( \callto{} \): \( \text{last}(\text{eval}(\tau', \sigma_0)) \).
    \item The \emph{evaluation} of a method call is the sequence of states resulting from evaluating the call segment prefix: \( \text{eval}(\tau_{\callto{}}, \sigma_0) \).
  \end{itemize}
\end{definition}

A flow \( \flowsto{\accpath{i.\overline{f}}}{\accpath{o.\overline{f'}}} \) between arbitrary access paths rooted at variables \( \var{i} \) and \( \var{o} \) is \emph{caller-visible} if \( \var{i}\) and \( \var{o}\) are in scope in the caller after the callee returns (i.e., \( \var{i}\in \sigma_E \) and \( \var{o}\in \sigma_E \)) where \( \sigma = \text{eval}(\tau_{\callto{}}, \sigma_0) \).
The \emph{summarized} taint flows of a method call are the \emph{transitive} caller-visible flows added by the callee.

Given a trace \( \tau \) with initial program state \( \sigma_0 \), call prefix \( \tau_{\callto{}} \preceq \tau \), call evaluation \( \overline{\sigma} = \text{eval}(\tau_{\callto{}}, \sigma_0) \) (see \cref{def:call-segment-prefix}), and call state \( \sigma_{\callto{}} \in \overline{\sigma} \), the summarized taint flows for a method call \( \callto{} \) is:
\[ \Psi_{\tau_{\callto{}}} \defeq {\{ \trflowsto{\accpath{i.\overline{f}}}{\accpath{o.\overline{f'}}} \in \text{last}(\overline{\sigma}) \mid \var{i} \in \incall{} , \var{o} \in \outcall{} \}} \, \setminus \, {\sigma_{{\callto{}}_T}} \]

The \emph{context-independent} taint flow model \( \Psi_m \) aggregates summarized flows across all calls to method \( m \) in all program traces:
\[ \Psi_m \defeq \bigcup_{\callto{} \in \text{call-sites}(m)} \bigcup_{{\tau \in \mathrm{T}} \mid {\tau_{\callto{}} \preceq \tau}} \Psi_{\tau_{\callto{}}} \]

A taint flow model \( \Phi_m \) is \emph{sound} if it over-approximates the summarized taint flows of \emph{any} call to method \( m \) in the program:
\[ \Phi_m \supseteq \Psi_m \]


%% file: sections/4_approach.tex
\section{Synthesizing And Checking Taint Flow Models}

Our approach generates taint flow models using LLMs and then checks the soundness of those models using a set of light-weight analyses. The soundness checking step may become as expensive as the full taint analysis. However, in most cases (see \Cref{sec:evaluation}), our approach, including the time it takes for the LLM to generate the model, is far more efficient than the constructive approach.

\subsection{LLM-based Model Synthesis}

Automatically generating models is necessary to leverage taint analysis tools on real programs. Manually writing taint flow models requires time, effort, and an understanding of the program being analyzed, the semantics of the language and the tool being used for taint analysis that a programmer may not have.

We believe that symbolic techniques are not a suitable tool to generate high-level taint flow models for the methods that are interesting to summarize. In our evaluation (\Cref{sec:evaluation}), we show that the naive approach of \emph{constructing} summaries by computing all flows within a method using taint analysis does not scale.
A generic program synthesis approach enumerating possible models would also likely not work well, because the scale of the problem is dictated by the precision required for a taint analysis, which we assume the LLM can choose well based on context.
For example, when taking into account field sensitivity, the number of possible models may grow too large for a simple method transforming a configuration file to a string, but the LLM would naturally not pick field sensitivity for a string transformation function, whose flow is likely a single edge from argument to output.
Since the requirement for a specific level of precision for a model is rather loose, it is difficult to predict with symbolic methods.



We propose to generate summaries using LLMs: they are usually good at understanding the high-level semantics of code, which includes understanding how data flows within a program. We build an LLM agent for writing taint flow models with a \emph{prompt} and a set of \emph{program analysis tools}.
Our prompt consists of a description of our taint flow model format, specific instructions to help with over-approximation (e.g., ``Account for all execution paths (branches, loops, error cases)''), and finally instructions on which program analysis tools should be used to generate a model. The tools that are available to the agent allow it to inspect the source code of the method, display its SSA form, access type information for specific variables and print aliasing information from a whole-program pointer analysis. The most important tool is the implementation of our soundness checking algorithm: the agent is able to check that the taint flow model it generated is sound, and fix it if necessary. However, since the agent is free to choose which tools to use, it may not always check models, or it may ignore the result of the soundness checking tool. Therefore, we always automatically check the soundness of the agent-generated model once it finishes. In \Cref{app:agent}, we provide the full prompt and list all the tools the agent can use.

\subsection{Incremental Soundness Checking Algorithm}

Our main contribution is the idea that a taint flow model can be checked sound instead of computed. We propose an algorithm that uses lightweight static analyses to prove that a given taint flow summary is a sound over-approximation of the inter-procedural taint flows within the method.
At a high-level, the algorithm tries to prove that the taint flows that are \emph{not} contained in the model \( \Phi_m \) do \emph{not} hold in the method \( m \) and all its transitive callees.


\subsubsection{Algorithm overview}

\begin{algorithm}
  \caption{Core soundness checking algorithm}\label{alg:soundness-checking}
  \KwIn{taint flow model $\Phi_m$}
  \KwOut{$\mathit{sound}$ if $\Phi_m$ is sound, $\mathit{unsound}$ otherwise}
  \SetKwFunction{FCheck}{CheckSoundness}
  \SetKwFunction{FDeduce}{DeduceCalleeSummaries}
  \SetKwProg{Fn}{function}{:}{}
  \Fn{\FCheck{$\Phi_m$}}{
    $\Phi_m^{\neg} \gets \Phi^\top_m \setminus \Phi_m$\;
    \ForEach{analysis $\analysis$\label{alg:line:analyses-start}}{
      \ForEach{$\nflowsto{\accpath{a}}{\accpath{b}} \in \Phi_m^{\neg}$}{
        \If{$m \models_{\analysis} \nflowsto{\accpath{a}}{\accpath{b}}$}{
          $\Phi_m^{\neg} \gets \Phi_m^{\neg} \setminus \{ \nflowsto{\accpath{a}}{\accpath{b}} \}$\;
        }
      }
    }\label{alg:line:analyses-end}
    \lIf{$\Phi_m^{\neg} = \emptyset$}{\KwRet $\mathit{sound}$}
    $\overline{\Phi_c} \gets$ \FDeduce{$\Phi_m^{\neg}$}\;\label{alg:line:deduce-start}
    \lIf{$\overline{\Phi_c} = \emptyset$}{\KwRet $\mathit{unsound}$}
    \ForEach{callee summary $\Phi_c \in \overline{\Phi_c}$}{
      \lIf{\FCheck{$\Phi_c$} $= \mathit{unsound}$}{\KwRet $\mathit{unsound}$}
    }\label{alg:line:deduce-end}
    \KwRet $\mathit{sound}$\;
  }
\end{algorithm}

\Cref{alg:soundness-checking} presents the core of our soundness checking algorithm. It requires a candidate taint flow model \( \Phi_m \).
It first computes the set of must-not-flows \( \Phi_m^{\neg} \), taint flows that cannot occur in \( m \) for \( \Phi_m \) to be sound, by removing all flows in \( \Phi_m \) from the most-general summary \( \Phi^\top_m \).
Then it uses a set of lightweight static analyses to prove each must-not-flow \( \nflowsto{\accpath{a}}{\accpath{b}} \) holds: \(m \models_{\analysis} \nflowsto{\accpath{a}}{\accpath{b}}\) indicates that analysis \(\analysis\) can prove that \(\accpath{a}\) never flows to \(\accpath{b}\) in \( m \). If a must-not-flow holds, it is removed from \( \Phi_m^{\neg} \). If there are any unproven must-not-flows remaining, it applies the \emph{intra-procedural} taint analysis to \( m \), deduces candidate models for the callees that are necessary for \( \Phi_m^{\neg} \) to hold, and then recursively checks their soundness.
The intra-procedural taint analysis yields the taint flow edges between the inputs and outputs of \(m\) and the inputs and outputs of the callees of \(m\), but not inside the callees. Given this information, we explain in \Cref{sec:deduce-callee-summaries} how we can deduce what the summary of each callee should be in order for the summary \( \Phi_m^{\neg} \) for \(m\) to be sound.
The key aspects of our approach are (1) the simplicity of the lightweight static analyses which are at most linear in the size of the program and (2) the ability to deduce the necessary summaries for the callees with \textsc{DeduceCalleeSummaries} given intra-procedural taint flows in \( m \).

\subsubsection{Most-general Summary}
The first step consists in determining the set of flows that are absent from the candidate summary relative to a \emph{most-general} summary \( \Phi_m^\top \), the top-level summary in the lattice of all possible summaries for \( m \).
Formally, the most-general summary \( \Phi^\top_m \subseteq \Pi \times \Pi \) is the set of all access path marks from the method's inputs to its outputs for \emph{all possible} calling contexts:
\[ \Phi^\top_m \defeq \bigcup_{(\var{i}, \var{o}) \in \inm{} \times \outm{}} \{ \flowsto{\accpath{i.\overline{f}}}{\accpath{o.\overline{f'}}} \mid (\accpath{i.\overline{f}}, \accpath{o.\overline{f'}}) \in \accpaths{i} \times \accpaths{o} \} \]
where \( \accpaths{i} \) is the set of all access paths rooted at input variable \( \var{i} \) and \( \accpaths{o} \) contains all access paths rooted at output variable \( \var{o} \), up to a constant length \( k \).

\begin{theorem}[Soundness]\label{thm:top-sound}
  The most general summary is sound; i.e., \( \Psi_m \subseteq \Phi^\top_m \).
\end{theorem}
\begin{proofsketch}
  The most-general summary is sound by construction as it contains all possible input-output access path flows in \( m \) for any calling context.
\end{proofsketch}

\subsubsection{Must-not-flows}
The \emph{must-not-flows} \( \Phi_m^{\neg} \) for the summary is a set containing the flows which \emph{must not} occur in the concrete summary \( \Psi_m \) for candidate model \( \Phi_m \) to be sound:
\[ \Phi_m^{\neg} \defeq \Phi^\top_m \setminus \Phi_m \]
A must-not-flow \( \nflowsto{\accpath{a}}{\accpath{b}} \in \Phi_m^{\neg} \) \emph{holds} in method \( m \), written \( m \models \nflowsto{\accpath{a}}{\accpath{b}} \), iff \( \flowsto{\accpath{a}}{\accpath{b}} \notin \Psi_m \).

\begin{theorem}[Soundness]\label{thm:mnf-sound}
  If \( m \models \nflowsto{\accpath{a}}{\accpath{b}} \) for all \( \nflowsto{\accpath{a}}{\accpath{b}} \in \Phi_m^{\neg} \), then \( \Phi_m \) is sound.
\end{theorem}
\begin{proofsketch}
  Assume \( m \models \nflowsto{\accpath{a}}{\accpath{b}} \) for all \( \nflowsto{\accpath{a}}{\accpath{b}} \in \Phi_m^{\neg} \), so \( \Psi_m \cap \Phi_m^{\neg} = \emptyset \).
  By \Cref{thm:top-sound}, \( \Psi_m \subseteq \Phi^\top_m \).
  Since \( \Phi_m^{\neg} \defeq \Phi^\top_m \setminus \Phi_m \), we have \( \Psi_m \subseteq \Phi^\top_m \subseteq \Phi_m \cup \Phi_m^{\neg} \), allowing us to conclude that \( \Psi_m \subseteq \Phi_m \).
  Therefore, \( \Phi_m \) is sound.
\end{proofsketch}

\subsection{Lightweight Analyses}\label{sec:static-analyses}

Each static analysis procedure \(\analysis\) in \Cref{alg:soundness-checking} takes as input a must-not-flow \( \nflowsto{\accpath{a}}{\accpath{b}} \) and returns true iff \( m \models \nflowsto{\accpath{a}}{\accpath{b}} \).
We instantiate our approach with a types, immutability and read analyses, driven by the following observations:
(1) if data from an input is never accessed (i.e., the input is unread) during the execution of the method, then taint can never flow \emph{from} this input; and (2) if data from an output is never modified (i.e., the output is immutable) during the execution of the method, then taint can never flow \emph{to} this output. We describe static analyses that check these intuitions.

\subsubsection{Types analysis}
Consider a simple type system where for any method call statement \( \call{} \), every argument variable \( \var{a} \in \mathbb{V} \) has a type of ``non-pointer-like'' if its value is unchanged after the method returns (i.e., \( a \) would be allocated on the stack if our language had a stack), or ``pointer-like'' otherwise.
Formally, an argument \( \var{a} \) is ``non-pointer-like'' if and only if:
\[ \denote{\callto{}}\sigma_H(\langle E(a), [] \rangle) = \denote{\callret{}}\sigma_H(\langle E(a), [] \rangle) \]

Recall that in our language, a parameter can also be an output of a method. If a variable \( p \) is a non-pointer-like parameter of a method \( m \), then its corresponding argument at any call to \( m \) is unchanged after \( m \) returns. Let the types analysis predicate \textsc{IsImmutableType}(\( \accpath{p.\overline{f}} \)) be true iff \( \accpath{p.\overline{f}} \) is an access path rooted at non-pointer-like parameter variable \( p \).

\begin{theorem}[Soundness]\label{thm:types-soundness}
  If \textsc{IsImmutableType}(\( \accpath{p.\overline{f}} \)) is true, then \( \forall \accpath{\pi} .\ m \models \nflowsto{\accpath{\pi}}{\accpath{p.\overline{f}}} \).
\end{theorem}
\begin{proofsketch}
  Since output variable \( \var{p} \) is not allocated on the heap, for any input access path \( \accpath{\pi} \), taint flow \( \flowsto{\accpath{\pi}}{\accpath{p.\varepsilon}} \) remains local to \( m \) and doesn't affect heap locations accessible to the caller.
  Therefore, \( \forall \accpath{\pi} .\ \flowsto{\accpath{\pi}}{\accpath{\pi.\varepsilon}} \notin \Psi_m \), giving us \( \forall \accpath{\pi} .\ m \models \nflowsto{\accpath{\pi}}{\accpath{p.\overline{f}}} \).
\end{proofsketch}

\subsubsection{Immutability analysis}\label{sec:immutability}
Given a must-not-flow \( \nflowsto{\accpath{a}}{\accpath{b}} \), the immutability analysis predicate \textsc{IsImmutablePtr}(\( \accpath{b} \)) is true iff (1) \( \accpath{b} \) and all of its aliases are never written to in \( m \) or any of its transitive callees; and (2) \( \accpath{b} \) is not a return value. Condition (2) is necessary because return values are always outputs of a method. Formally, for every memory write statement of the form \( \store{} \) in \( m \) and its transitive callees,
\[ \textsc{IsImmutablePtr}(b) \iff (\ptranalysis(\accpath{x.f}) \cap \ptranalysis(\accpath{b}) = \emptyset) \land \accpath{b} \text{ is not returned} \]

\begin{theorem}[Soundness]\label{thm:immutability-soundness}
  If \textsc{IsImmutablePtr}(\( \accpath{b} \)) is true, then \( \forall \accpath{a} .\ m \models \nflowsto{\accpath{a}}{\accpath{b}} \).
\end{theorem}
\begin{proofsketch}
  If \( \accpath{b} \) is immutable, then (1) for every store statement \( \store{} \) in \( m \) and its transitive callees, we have \( \ptranalysis(\accpath{x.f}) \cap \ptranalysis(\accpath{b}) = \emptyset \), and (2) \( \accpath{b} \) is never returned to the caller.
  For condition (1), for the taint flow equation for the store statement to add a flow \( \flowsto{\accpath{a}}{\accpath{b}} \in \Psi_m \), we need \( \accpath{b} \in \ptranalysis(\accpath{x.f}) \).
  This requires \( \ptranalysis(\accpath{b}) \cap \ptranalysis(\accpath{x.f}) \neq \emptyset \), which contradicts the immutability predicate.
  For condition (2), for the taint flow equation for the return statement to add the flow, we need \( \accpath{b} \) to be a return value, which contradicts the immutability predicate.
  Therefore, \( \forall i .\ \flowsto{\accpath{a}}{\accpath{b}} \notin \Psi_m \), giving us \( \forall \accpath{a} .\ m \models \nflowsto{\accpath{a}}{\accpath{b}} \).
\end{proofsketch}

\subsubsection{Read analysis}
The read analysis is essentially the inverse of the immutability analysis.
Given a must-not-flow input access path \( \accpath{a} \), the read analysis predicate \textsc{IsUnread}(\( \accpath{a} \)) is true iff \( \accpath{a} \) and all of its aliases are never read from in \( m \) or any of its transitive callees. Formally, for every memory load statement of the form \( \load{} \) in \( m \) and its transitive callees,
\[ \textsc{IsUnread}(\accpath{a}) \iff \ptranalysis(\accpath{y.f}) \cap \ptranalysis(\accpath{a}) = \emptyset \]

\begin{theorem}[Soundness]\label{thm:read-soundness}
  If \textsc{IsUnread}(\( \accpath{a} \)) is true, then \( \forall \accpath{b} .\ m \models \nflowsto{\accpath{a}}{\accpath{b}} \).
\end{theorem}
\begin{proofsketch}
  If \( \accpath{a} \) is unread, then for every load statement \( \load{} \) in \( m \) and its transitive callees, we have \( \ptranalysis(\accpath{y.f}) \cap \ptranalysis(\accpath{a}) = \emptyset \).
  For the taint flow equation for the load statement to add a flow \( \flowsto{\accpath{a}}{\accpath{y.f}} \in \Psi_m \), we need \( \accpath{a} \in \ptranalysis(\accpath{y.f}) \).
  This requires \( \ptranalysis(\accpath{a}) \cap \ptranalysis(\accpath{y.f}) \neq \emptyset \), which contradicts the unread condition.
  Therefore, \( \forall \accpath{b} .\ \flowsto{\accpath{a}}{\accpath{b}} \notin \Psi_m \), giving us \( \forall \accpath{b} .\ m \models \nflowsto{\accpath{a}}{\accpath{b}} \).
\end{proofsketch}

\subsection{Deducing Callee Summaries}\label{sec:deduce-callee-summaries}


We still have remaining must-not-flows that we need to prove hold in \( m \), the method under analysis, but we do not want to discharge them using the only static analysis method we have left: an \emph{inter}-procedural taint analysis, as this would be as inefficient as the constructive approach.
Reps et al.~\cite{repsPreciseInterproceduralDataflow1995} demonstrate that inter-procedural taint analysis is a graph reachability problem. We use this intuition to express our maximally-general but sound taint flow summary for \( m \) as an inter-procedural taint flow graph.
To avoid performing an inter-procedural taint analysis, we instead compute the \emph{intra}-procedural taint flows in \( m \) and then \emph{deduce} the summarized taint flows of each callee of \( m \) such that the unproven must-not-flows hold.
This gives us an incomplete inter-procedural taint flow graph \( G = (V, E) \) for \( m \), as it only contains intra-procedural edges. The graph has implied inter-procedural edges from call sites to callee inputs/outputs, but is missing intra-procedural edges from callee inputs to outputs.
To complete this graph, we translate it into a Max-SAT problem to deduce the \emph{least precise} (maximally-general) intra-procedural edges for the callees of \( m \) that satisfy the unproven must-not-flows in \( \Phi_m^{\neg} \).

We now describe the implementation of \textsc{DeduceCalleeSummaries} in \Cref{alg:soundness-checking} in more detail.
We require a sound field-sensitive intra-procedural static taint analysis \( \mathcal{T} \) which over-approximates the concrete taint flows in a program.
We initialize \( \mathcal{T} \) for \( m \) by tainting every input access path in \( \inm{} \).
Since \( \inm{} \subseteq \mathbb{V} \) but our taint flow functions (see \Cref{sec:taint-semantics}) operate on access paths, we must enumerate all the access paths for each input up to a maximum length \( k \), where \( k \) is determined by the longest access path in the candidate model \( \Phi_m \). If \( k = 0 \) then the taint analysis can be field-insensitive as the model only has field-insensitive flows.

The intra-procedural taint analysis proceeds as described by the flow functions in \Cref{sec:taint-semantics}, with key modifications to avoid analyzing intra-procedural flows in the callees.
When a taint flow reaches a call statement, instead of descending into the callee, the analysis adds inter-procedural taint flows for each tainted call site input (argument) to the corresponding callee input (parameter).
It then assumes the most-general summary for the callee, where all tainted callee inputs flow to all callee outputs.
If a callee output is tainted, it adds inter-procedural taint flows from the callee output to the corresponding call site output.
This special handling of callees is important because otherwise, must-not-flows along callee outputs would be vacuously satisfied, which is unsound.
The taint analysis stops when it has computed all possible intra-procedural taint flows from inputs in \( \inm{} \) to outputs in \( \outm{} \).

The graph vertices in \( V \) represent access paths at a calling context which include:
\begin{enumerate}
  \item Inputs and outputs of \( m \).
  \item Inputs and outputs of each call site in \( m \) (e.g., arguments and return values, respectively).
  \item Inputs and outputs of each callee. Since \( m \) can call the same callee \( c \) multiple times with different inputs, resulting in different \emph{calling contexts}, we differentiate callee vertices from different call statements \( s \) and \( s' \) by appending the statement to the callee node. E.g., for a callee parameter \( p \) from call site \( s = {\downarrow} c(a) \), we have node \( p.\varepsilon@s \), which is distinct from \( p.\varepsilon@s' \) representing the same parameter at call site \( s' = {\downarrow} c(a') \).
\end{enumerate}

Adding the most-general flows for each callee allows the \emph{intra}-procedural taint analysis to compute \emph{all possible} taint flows from inputs of \( m \) to its outputs, without analyzing the callees.
This is all the information we need to connect all the vertices in \( G \) with edges: every known edge comes from the intra-procedural taint analysis and every unknown edge is one of the most-general callee flows.

We compute \( G \) as a least fixpoint, giving us the set of vertices reachable from the input access paths. The initial set of vertices are the enumerated input access paths of \( m \) with an empty calling context. We then repeatedly pick a vertex whose outgoing edges have not yet been computed and add them, along with any newly reachable vertices, until every reachable vertex is in \( G \). The outgoing edges include:
\begin{enumerate}
  \item \emph{Known} intra-procedural edges of \( m \): edges between nodes in \( m \) from the intra-procedural taint analysis. This also includes edges reachable from call site output nodes.
  \item \emph{Known} inter-procedural edges of \( m \): edges from call site input nodes to their corresponding callee input nodes, and edges from callee output nodes to their corresponding call site output nodes. These edges are known because if taint flows to a call site input in \( m \), it \emph{must} flow to the corresponding callee input, and likewise for outputs.
  \item \emph{Unknown} intra-procedural edges of a callee: edges from callee input nodes to callee output nodes. These edges form the callee summary to be deduced.
\end{enumerate}
The computation terminates even if \( m \) is recursive because \( V \) is a finite set: access paths are bounded and a vertex's calling context is only the immediate call site, so \( |V| \) is bounded by the number of call sites in \( m \) rather than the recursion depth.

Once we have our complete inter-procedural graph \( G \), we must deduce taint flow summaries for the callees that \emph{satisfy} the unproven must-not-flows from the lightweight static analyses.
Recall that \Cref{alg:soundness-checking} \emph{recursively} proves the soundness of each deduced callee summary.
We want to deduce the \emph{least precise} callee summaries (most taint flow edges) to minimize the number of must-not-flow edges we need to prove hold in the callees.
We do this by encoding the reachable vertices of \( G \) and the must-not-flows as a Max-SAT problem.

Given a set of hard and soft constraints as logical formulas, a Max-SAT solver maximizes the number of soft constraints that can be true while satisfying the hard constraints.
Our Max-SAT domain has two kinds of Boolean variables: a variable \( \flowsto{u}{v} \) for each edge \( (u, v) \in E \), which holds iff taint may flow from \( u \) to \( v \) in one step; and a variable \( \reach{u_0}{v} \) for each input access path vertex \( u_0 \) and each vertex \( v \) reachable from it in \( G \), which holds iff taint may flow from \( u_0 \) to \( v \) along a chain of edges.

We assert every known edge of \( G \) as a hard constraint.
A must-not-flow forbids taint from an input to \emph{transitively} reach an output. Blocking a must-not-flow therefore requires asserting the absence of a path along \( G \) rather than of an edge.
We thus encode the reachability relation on vertices \( \reach{\cdot}{\cdot} \) inductively: every input access path vertex reaches itself, and if \( u_0 \) reaches \( u \) and the edge \( \flowsto{u}{v} \) holds, then \( u_0 \) reaches \( v \).
Our formulation directly enforces context-independence of callee summaries: it names each unknown edge by its callee and its endpoints' access paths rather than by the call site that observed it, so all of its call sites share a single variable.
Finally, we maximize the unknown callee edges via soft constraints.
Maximizing unknown edges yields the least precise summaries because more taint flow edges mean fewer restrictions, reducing the static analysis burden for proving must-not-flows in callees.

We provide our Max-SAT formulation in \Cref{fig:maxsat-formulation}.
An unsatisfiable model is not necessarily an error---e.g., it can occur when \( \Phi_m^{\neg} \) contradicts the intra-procedural analysis results alone. In such cases, the algorithm returns the empty set, indicating that no valid callee summaries exist that satisfy the soundness requirements.

\begin{figure}
  \centering
  \setlength{\tabcolsep}{3pt}
  \begin{tabular}{@{}l l l l@{}}
    \textbf{Variables:}
    & \multicolumn{3}{l}{\( \flowsto{u}{v} \) for each edge \( (u, v) \in E \)} \\
    & \multicolumn{3}{l}{\( \reach{u}{v} \) for each \( u \in V(\accpaths{\inm{}}) \) and each \( v \in V \) reachable from \( u \) in \( G \)} \\[0.5em]
    \textbf{Hard:}
    & \( \bigwedge_{(u,v) \in E \setminus \mathit{soft}(E)} \)
    & \( \flowsto{u}{v} \)
    & \emph{(known edges)} \\
    & \( \bigwedge_{u_0 \in V(\accpaths{\inm{}})} \)
    & \( \reach{u_0}{u_0} \)
    & \emph{(base case)} \\
    & \( \bigwedge_{\substack{u_0 \in V(\accpaths{\inm{}}) \\ (u,v) \in E}} \)
    & \( \reach{u_0}{u} \land \flowsto{u}{v} \implies \reach{u_0}{v} \)
    & \emph{(inductive step)} \\
    & \( \bigwedge_{\substack{\nflowsto{\accpath{a}}{\accpath{b}} \in \Phi_m^{\neg} \\ u \in V(\accpath{a}),\, v \in V(\accpath{b})}} \)
    & \( \neg \reach{u}{v} \)
    & \emph{(must-not-flows)} \\[0.5em]
    \textbf{Soft:}
    & \( \bigwedge_{(u,v) \in \mathit{soft}(E)} \)
    & \( \flowsto{u}{v} \)
    & \emph{(maximize callee edges)}
  \end{tabular}
    \caption{MaxSAT formulation for deducing callee summaries. \( G = (V, E) \) is the inter-procedural taint flow graph; each vertex pairs an access path with a calling context. \( V(\accpath{a}) \) denotes the vertices with an empty calling context whose access path overlaps \( \accpath{a} \), extended pointwise to sets of access paths, so \( V(\accpaths{\inm{}}) \) are the input access path vertices.}
  \label{fig:maxsat-formulation}
\end{figure}

There may be multiple Max-SAT solutions which satisfy the model, as the soft constraints have the same weight. For example, there may be two maximal deduced summaries for a callee \( c \) that have the same number of taint flows. For each solution, the algorithm converts the satisfying assignments of the edge variables to summary edges for each callee and returns them.
We then \emph{recursively} check the soundness of a deduced summary for each callee (if there are multiple, we pick one), concluding that \( \Phi_m \) is sound iff the summaries for \emph{all} callees of \( m \) are sound.

We add an extra case to \Cref{alg:soundness-checking} to ensure termination when checking the soundness of recursive methods. If it is unable to prove the must-not-flows using the static analyses for a callee whose summary it deduced before, we conservatively consider the method's model to be unsound.

\begin{theorem}[Soundness]\label{thm:deduce-sound}
  If we deduce callee summaries \( \overline{\Phi_c} \) and all the deduced callee summaries are sound, then \( \Phi_m \) is sound.
\end{theorem}
\begin{proofsketch}
  Assume each deduced callee summary \( \Phi_c \in \overline{\Phi_c} \) is sound; i.e., \( \Phi_c \supseteq \Psi_c \).
  We show \( m \models \nflowsto{\accpath{a}}{\accpath{b}} \) for all \( \nflowsto{\accpath{a}}{\accpath{b}} \in \Phi_m^{\neg} \), which by \Cref{thm:mnf-sound} implies \( \Phi_m \) is sound.

  The intra-procedural taint analysis assumes the most-general summary \( \Phi^\top_c \) for each callee \( c \). By \Cref{thm:top-sound}, \( \Phi^\top_c \supseteq \Psi_c \). Since the intra-procedural analysis is sound and \( \Phi^\top_c \) over-approximates all callee flows, any concrete summarized flow in \( \Psi_m \) consists of (1) intra-procedural edges in \( m \), which are over-approximated by the \( \mathcal{T} \), and (2) callee input-to-output edges, which are contained in \( \Phi^\top_c \). The flow thus begins at an input access path vertex and consists of edges that the construction of \( G \) adds. Since \( G \) is the least graph closed under those edges, induction on the length of the flow shows that every vertex it visits belongs to \( V \) and that consecutive vertices are joined by an edge in \( E \): every concrete summarized flow in \( \Psi_m \) has a corresponding path in \( G \).

  Suppose for contradiction that \( \flowsto{\accpath{a}}{\accpath{b}} \in \Psi_m \) for some \( \nflowsto{\accpath{a}}{\accpath{b}} \in \Phi_m^{\neg} \). Then there is a witnessing path in \( G \) from some \( u \in V(\accpath{a}) \) to some \( v \in V(\accpath{b}) \). Consider a satisfying model that the solver returns. Every edge along the witnessing path is true in the model: the known edges because they are asserted as hard constraints; and the unknown edges because the concrete flow's step through each callee is a flow in \( \Psi_c \subseteq \Phi_c \), and \( \Phi_c \) consists of exactly the callee flows whose edges the model assigns true. Applying the reachability implication once per edge along the path, by induction on its length, forces \( \reach{u}{v} \) to be true. However, the witnessing path shows that \( v \) is reachable from \( u \) in \( G \), so the Max-SAT encoding contains the hard must-not-flow constraint \( \neg \reach{u}{v} \), which the model returned from the solver satisfies---a contradiction. Therefore, no such concrete summarized flow in \( \Psi_m \) can exist.

  Therefore \( m \models \nflowsto{\accpath{a}}{\accpath{b}} \) for all \( \nflowsto{\accpath{a}}{\accpath{b}} \in \Phi_m^{\neg} \), and \( \Phi_m \) is sound by \Cref{thm:mnf-sound}.
\end{proofsketch}

\subsection{Efficiency}

\Cref{alg:soundness-checking} requires a candidate taint model \( \Phi_m \), a pointer analysis \( \ptranalysis \), and a call graph.
We assume that the compiler provides the program's intermediate representation and that the LLM generates \( \Phi_m \). We discuss our empirical results for LLM model generation time in \Cref{sec:evaluation}.
Let \( |m| \) denote the number of statements in method \( m \) and \( C^\ast \) the set of transitive callees of \( m \) reachable from the call graph.
Let \( T_\ptranalysis \) denote the time to compute pointer analysis results \( \ptranalysis \). For simplicity, we assume that once \( \ptranalysis \) is computed, may-alias queries take constant time. If we use an on-demand pointer analysis for \( \ptranalysis \), the time to compute each may-alias query for every statement in \( m \) is also \( T_\ptranalysis \).

We use \( \Omega \) to state the asymptotic \emph{lower} bounds for our approach to better compare its efficiency with the constructive approach. For completeness we also provide the \( O \) asymptotic higher bounds, using \( \Theta \) when \( O(\cdot) = \Omega(\cdot) \).

The lightweight static analyses in lines \ref{alg:line:analyses-start}--\ref{alg:line:analyses-end} of \Cref{alg:soundness-checking} run in terms of \( |m| \) and \( |C^\ast| \):
\begin{itemize}
  \item The most-general summary computation and types analysis run in constant time with respect to \( |m| \) as they only need to analyze the parameters of \( m \), not its statements.
  \item The immutability and read analyses run in \( \Theta(|m| + \sum_{c \in C^\ast} |c|) \) time, analyzing all statements in \( m \) and its transitive callees for each unproven must-not-flow.
\end{itemize}

The callee summary deduction (lines \ref{alg:line:deduce-start}--\ref{alg:line:deduce-end}) depends on the choice of taint analysis, which uses the pointer analysis \( \ptranalysis \).  Let \( T_{\text{intra}}(m) \) be the time for the intra-procedural taint analysis of method \( m \), and let \( G_m = (V_m, E_m) \) be the taint flow graph built from its result, with \( I_m \subseteq V_m \) the input access path vertices. Constructing \( G_m \) expands each vertex exactly once, so it takes \( \Theta(|V_m| + |E_m|) \) time, assuming each outgoing edge is computed in constant time.

The Max-SAT problem is NP-complete in general and its run time scales with the size of its domain and the number of clauses it must solve for~\cite{halabyComputationalComplexityMaxSAT2016}. The number of clauses is dominated by the reachability constraints, which contribute one clause per input access path vertex and edge reachable from it, giving \( O(|I_m| \cdot |E_m|) \). The known-edge and soft constraints contribute \( O(|E_m|) \). The must-not-flow constraints are no more numerous than the reachability constraints: each blocks a pair whose target is reachable from an input access path vertex, and every such target is reached by an edge already counted above. This also dominates the cost of constructing \( G_m \), so we omit the construction term below. Enumerating the co-optimal models requires a further solver call per model, which we bound by a constant.

Since the algorithm is recursive, it runs at most \( 1 + |C^\ast| \) times. The total \emph{worst-case} complexity of our soundness checking algorithm is:
\[ O\left( T_\ptranalysis + \sum_{c \in C^\ast} \left( |c| + T_{\text{intra}}(c) + \text{Max-SAT}(|I_c| \cdot |E_c|) \right) \right) \]

The key efficiency advantage over the constructive approach is that our algorithm does not need to analyze every transitive callee in the best case: it only recurses into callees whose deduced summaries have unproven must-not-flows.
Therefore, we argue that the \( \Omega \) complexity comparison is more useful.
Our approach pays an additional Max-SAT cost but only for the subset of callees that reach the callee summary deduction step. We show empirically that this, along with the low number of Max-SAT constraints, leads to significant speedups over the constructive approach in practice (\Cref{sec:evaluation}).
In the \emph{best} case, our approach has complexity:
\[ \Omega\left(  T_\ptranalysis + |m| + T_{\text{intra}}(m) + \text{Max-SAT}(|I_m| \cdot |E_m|) \right) \]

For comparison, the constructive approach (classical inter-procedural taint analysis) has worst \emph{and} best-case complexity:
\[ \Theta\left( T_\ptranalysis + \sum_{c \in C^\ast} T_{\text{intra}}(c) \right) \]
as it must run the intra-procedural taint analysis on \emph{every} transitive callee to compute the summary of \( m \).


%% file: sections/5_instantiation.tex
\section{Instantiation for Go}
\label{sec:instantiation}

We instantiate our approach for the Go programming language. As an imperative programming language that has pointers and methods, Go resembles the simple language we defined in \Cref{sec:simple-language}. However, as a general-purpose modern programming language, it is too complex to fully specify here.
While we provide taint flow equations for all statements in our simple language, Go has language features that a sound taint analysis cannot analyze efficiently and precisely.
These features include, but are not limited to, unsafe pointer manipulation, reflection, and concurrency.
To our knowledge, no static taint analysis exists for Go that models these features.
We instantiate our approach's taint analysis with the \textsc{Argot}~\cite{argot} (\underline{A}utomated \underline{R}easoning \underline{Go} \underline{T}ools) static analysis.

\subsection{Taint Flow Model}
\label{sec:instantiation-model}

Go has global variables and closures which complicates taint flow summaries. Parameters are passed by-value, but values in Go can be pointers. We extend the method input set \( \inm{} \) to include all global variables in the package containing \( m \) and all free variables if \( m \) is a closure. The output set \( \outm{} \) includes all variables in \( \inm{} \) along with the return values.

Functions in Go can also be higher-order. Methods and functions are equivalent in Go: a method is a function with its object receiver as the first parameter. We define a higher-order function as any function that has an input or output access path which resolves to a function. Since our taint model format does not support higher-order functions, we can only prove the soundness of taint models for single-order functions. For example, consider a closure \code{g := func() \{ *y = *x \}} that when executed, will produce a taint flow from bound variable \code{x} to \code{y}. This closure can be passed directly as an argument to a function \code{f}, or passed indirectly inside of an object (we use object and the more common \code{struct} interchangeably) \code{o := struct\{h func()\}\{h: g\}} that contains the closure. We now cannot model the taint flows inside \code{func f(g func()) \{ ... \}} because if \code{f} ever executes \code{g}, then there will be a taint flow that is not contained in the \emph{explicit} inputs or outputs of \code{f}.

Our \emph{checkable} taint model format does not encode flows to or from global variables or bound/free variables, as these constructs do not appear in the type signature of a Go function.
Therefore, our soundness checking algorithm will declare a model to be unsound if the function being modeled is higher-order or has any of these unsupported inputs and outputs. \textsc{Argot}'s taint analysis fully supports analyzing these features; it is only a limitation of our \emph{model} format.

We \emph{can} check the soundness of any callee summary that is a closure if its inputs and outputs are fully resolved (i.e., its free variables are bound to either an input of \( m \) or a value allocated in the scope of \( m \)). However, if \( m \) or any of its callees reads from or writes to a global variable, we conservatively consider the model to be unsound.

Go supports dynamic dispatch via interfaces. An interface is a set of methods which an object can implement.
If an input or output of a taint model has an interface type, it must be field-insensitive, because its fields may depend on the concrete type the program instantiates the interface with at run-time.

\textsc{Argot} supports user-provided taint models for interfaces as well as individual functions.
Our approach checks that a model is sound for an interface method by checking that it is sound for every concrete implementation.
Since interface models must be independent of their implementations, we do not support field-sensitive inputs or outputs.
For example, the \code{fmt.Stringer} interface specifies a single method \code{func (T) String() string}; every type \code{T} that has a \code{func (T) String() string} method implements the \code{fmt.Stringer} interface.
To check the soundness of a given model for the \code{fmt.Stringer} interface \code{String} method, we check the soundness of the model for every \code{String} method of each object that implements the interface in the application. As a result, our soundness checking approach is only sound with respect to the application.

In summary, we support checking the soundness of the following model types:
\begin{enumerate}
  \item \textbf{Function:} parameter inputs to parameter and return value outputs
  \item \textbf{Method:} object receiver (field-sensitive) and parameter inputs to object receiver (field-sensitive), parameter, and return value outputs
  \item \textbf{Interface method:} object receiver (field-insensitive) and parameter inputs to object receiver (field-insensitive), parameter, and return value outputs
\end{enumerate}

Our models also do not support method inputs and outputs that come from outside the application, such as standard input (e.g., via \code{fmt.Scanf}). We partially mitigate this limitation by considering a model to be unsound if any of its callees are sources.

\subsection{Most-General Summary}

The most-general summary is unsound in the presence of unsafe pointer manipulations. Consider for example a method with a parameter \code{x} of type \code{int32}. If this method obtains the raw pointer address of \code{x}, adds an arbitrary offset to the address, and then dereferences the value of the offset pointer, the method's environment now contains a reference to potentially any object in the program. This new access path may not be in the input set \( \inm{} \), making the most-general summary unsound. The input set cannot represent this new access path.

We assume that the method's execution is data-race-free such that heap reads and writes are linearizable, as defined in the Go memory model~\cite{go-memory-model}. If there is a data race, an instruction may read from memory via an access path not in \( \inm{} \) or write to memory via an access path not in \( \outm{} \), making the most-general summary unsound. We could discharge this assumption by using a dedicated data race analysis; however, to our knowledge, none yet exist for Go that can scale to large programs.

\begin{theorem}
  If the program is data-race-free and \( m \) does not unsafely manipulate memory, then \( \Phi_m^\top \) is sound.
\end{theorem}
\begin{proofsketch}
  We assume that the only other statements in \( m \) that can violate the soundness of \( \Phi_m^\top \) are reflection and concurrency operations.

  No operation in the Go reflection API can add new access paths to \( \inm{} \). Therefore, \( \Phi_m^\top \) is sound for all uses of the reflection API.

  Suppose for contradiction that for any statement \( s \) in \( m \), any other statement \( s' \) in the program can execute before, at the same time as, or after \( s \). Due to the scoping rules of Go, \( s' \) cannot add a taint flow from any access path referenced in \( s \) that is not in \( \inm{} \). Therefore, \( \Phi_m^\top \) is sound in the presence of concurrency.
\end{proofsketch}

\begin{theorem}
  Under the assumptions of analysis \( \mathcal{A}\), if \( m \models_{\mathcal{A}} \nflowsto{\accpath{a}}{\accpath{b}} \) for all \( \nflowsto{\accpath{a}}{\accpath{b}} \in \Phi_m^{\neg} \), then \( \Phi_m \) is sound.
\end{theorem}

\subsection{Lightweight Static Analyses}

\subsubsection{Types analysis}

In Go, free variables, global variables, return values, and pointer-like parameters are accessible to the caller of a function. We consider a parameter object to be pointer-like if it or any of its elements/fields have a pointer-like type. We consider slices, maps, channels, interfaces (including the empty interface type \code{interface\{\}} or \code{any}), and pointers to be pointer-like types. Go is statically typed; therefore, every object in a Go program has a type which is either pointer-like or non-pointer-like according to our definition. Let \textsc{IsImmutableType}(\( \accpath{p.\overline{f}} \)) be true iff \( \var{p} \) is a non-pointer-like parameter in \( \outm{} \).

\begin{theorem}[Soundness]
  Assuming the program is data-race-free and \( m \) does not unsafely manipulate memory, if \textsc{IsImmutableType}(\( \accpath{p.\overline{f}} \)) is true, then \( \forall \accpath{\pi} .\ m \models \nflowsto{\accpath{\pi}}{\accpath{p.\overline{f}}} \).
\end{theorem}
\begin{proofsketch}
  The Go type system is sound under our assumptions. Therefore, if \textsc{IsImmutableType}(\( \accpath{p.\overline{f}} \)) is true, then output variable \( \var{p} \) must not be allocated on the heap. The rest of the proof follows from the proof of \Cref{thm:types-soundness}.
\end{proofsketch}

\subsubsection{Immutability and read analyses}

We use Go's first-party pointer analysis~\cite{go-pointer-analysis}: a flow-insensitive, context-insensitive, field-sensitive, and \emph{object-sensitive} implementation of Andersen's analysis~\cite{andersenProgramAnalysisSpecialization1994}.
Object sensitivity~\cite{milanovaParameterizedObjectSensitivity2002,smaragdakisObjectSensitivity2011} uses an object's allocation site as the context, resulting in increased precision.

Our immutability and read analyses for Go are unsound in the presence of reflection because our pointer analysis cannot soundly model the full reflection API for Go. A flow-insensitive pointer analysis is sound in the presence of concurrency~\cite{deDataflowAnalysisDataraceFree2011}. If we use a flow-sensitive pointer analysis, gaining more precision, the immutability and read analyses would be unsound in the presence of concurrency.

Go has constant values, which our language in \Cref{sec:simple-language} does not. Therefore, we can make our immutable analysis more precise for the return value special case.
We extend the immutability analysis predicate \textsc{IsImmutablePtr} from \Cref{sec:immutability} to treat \emph{constant} return values as immutable.

\begin{theorem}[Soundness]
  Assuming the program is data-race-free, \( m \) does not unsafely manipulate memory or use reflection, and \textsc{IsImmutablePtr}(\( \accpath{b} \)) is true, then \( \forall \accpath{a} .\ m \models \nflowsto{\accpath{a}}{\accpath{b}} \).
\end{theorem}
\begin{proofsketch}
  If our assumptions hold, then pointer analysis \( \mathcal{P} \) is sound. The proof follows from the proof of \Cref{thm:immutability-soundness}.
\end{proofsketch}

\begin{theorem}[Soundness]
  Assuming the program is data-race-free, \( m \) does not unsafely manipulate memory or use reflection, and \textsc{IsUnread}(\( \accpath{a} \)) is true, then \( \forall \accpath{b} .\ m \models \nflowsto{\accpath{a}}{\accpath{b}} \).
\end{theorem}
\begin{proofsketch}
  If our assumptions hold, then pointer analysis \( \mathcal{P} \) is sound. The proof follows from the proof of \Cref{thm:read-soundness}.
\end{proofsketch}

The benefit of our approach is that a flow-insensitive pointer analysis can be precise enough to prove the soundness of a given taint flow model. If not, we could extend our approach to automatically use a flow-sensitive implementation when needed. Whereas for the constructive approach, the taint analysis user would need to manually decide which pointer analysis to use to compute a precise model.

\subsection{Deducing Callee Summaries}

\subsubsection{Taint analysis}

\textsc{Argot}'s taint analysis is flow-sensitive, context-sensitive, and field-sensitive. The taint analysis uses the same pointer analysis as our immutability and read analyses to resolve aliases.
It is sound under some key assumptions; namely that unsafe pointer manipulation, reflection, and concurrency do not affect taint flows. While it does not check these assumptions, it does report when these features are used in a method it is analyzing. We could use a thread escape analysis to ensure that all reads and writes within the method operate on access paths that are local to the current thread, which would make the taint analysis sound in the presence of concurrency.

The taint analysis does not yet soundly compute taint flows in the presence of some Go-specific control flow constructs such as \code{panic}/\code{recover} statements (which are similar to exceptions) and unbounded \code{defer} statements (which execute before the method returns). Our approach could automatically use a flow-insensitive analysis to be sound in these cases, at the expense of precision. The taint analysis inherits all of the soundness limitations of the Go type system and pointer analysis: unsafe memory manipulations, data races, and reflection.

\subsubsection{Callee models}

\begin{wrapfigure}{r}{.5\textwidth}
  \begin{lstlisting}
func nestedClosures(x, y *int) *int {
  bv := *y
  outer := func(z *int) *int {
    inner := func() *int {
      return z
    }
    res := *inner() + bv
    return &res
  }
  return outer(x)
}
  \end{lstlisting}
  \caption{Closure summarization}
  \label{fig:ex-closures}
\end{wrapfigure}

While our approach does not support checking the soundness of methods with free variables, it does handle checking the soundness of methods that \emph{call} closures. Consider \Cref{fig:ex-closures}, which defines a function \code{nestedClosures} that contains two closures: \code{outer} and \code{inner}. Given a model \tmodel{x $\tto$ return, y $\tto$ return}, our approach deduces the following callee summaries: \mcode{outer:} \tmodel{z $\tto$ bv, z $\tto$ return, bv $\tto$ return} and \code{inner:} \tmodel{z $\tto$ return}. In \code{outer}, \code{bv} is a free variable which is bound to the corresponding variable in \code{nestedClosures}; and in \code{inner}, \code{z} is a free variable which is bound to parameter \code{z} of \code{outer}.

\subsection{Limitations}\label{sec:limitations}

Our approach can only check the soundness of taint flow models for methods that meet the restrictions we describe in \Cref{sec:instantiation-model}. Due to dynamic dispatch, which is pervasive in Go programs, our soundness proofs are relative to the application.

Our soundness proof sketches are relative to our interpretation of the Go programming language specification~\cite{go-spec}. To the best of our knowledge, Go does not have formal semantics. Go may have additional semantics that we do not consider which could compromise the correctness of our soundness checking algorithm. If we misinterpret the semantics of Go, our soundness checking algorithm may incorrectly claim that an unsound model is sound.

We assume that all instructions in the execution of a modeled method are free of data races and that the program's call graph is sound. We also assume that there are no free variables in any input or output of the methods we model. These are the only assumptions that we do not validate in the instantiation of our approach.

Modeling and analyzing \emph{implicit}~\cite{sabelfeldLanguagebasedInformationflowSecurity2003} taint flows to prove non-interference properties (e.g., the absence of control-flow-based taint flows) is out of scope for our approach.


%% file: sections/6_evaluation.tex
\section{Evaluation}
\label{sec:evaluation}

Taint analyses analyze real-world programs for violations of certain properties specified as data flow reachability problems. For example, one such property is \emph{data from a user input (source) must never reach a log (sink)}. We evaluate the soundness, precision, and efficiency of our approach with respect to a baseline traditional inter-procedural taint analysis. In particular, given a taint flow property for a program, we analyze the property using the baseline taint analysis, which computes taint flows on-demand by only computing the intra-procedural taint flows of methods that are reachable from a source. Afterwards, we identify interesting methods; i.e., methods that were expensive to intra-procedurally analyze, made the analysis unsound (e.g., reflection), or caused an explosion in the size of the search space (e.g., calling an interface method with hundreds of implementations). Each interesting method participates in taint flow propagation for a given property.

Our approach uses an LLM agent to generate taint flow models for each interesting method in the program and checks its soundness. We then run the same inter-procedural taint analysis as the baseline, but at call sites to interesting methods, instead of intra-procedurally analyzing the callee, we use our checked-sound model of the callee to propagate taint flow.

In a practical setting, taint analysis users must re-run the analysis after each change to the program to ensure that the program maintains the specified properties. For the baseline taint analysis, this means re-computing the intra-procedural taint flows of each method reachable from a source. However, when applying our approach, a user only needs to use the LLM to generate the taint flow models \emph{once}, but must check the soundness of each LLM-generated model after every program change.

Our evaluation answers the following research questions:
\begin{itemize}
    \rqitem{rq:checker-precision}{How \emph{precise} is the soundness checker? I.e., how often does it classify a most-precise sound taint flow model as unsound?}
    \rqitem{rq:checker-efficiency}{How \emph{efficient} is the soundness checker, compared to the constructive approach?}
    \rqitem{rq:checker-ablation}{How useful are the light-weight analyses we instantiate the soundness checker with?}
    \rqitem{rq:llm-effectiveness}{How \emph{sound} and \emph{precise} are the taint flow models that the LLM agent generates?}
    \rqitem{rq:workflow-efficiency}{Does using the checked-sound LLM-generated models make the taint analysis tractable for properties where it would otherwise time out?}
\end{itemize}

\begin{table}[t]
  \caption{Go repositories used to benchmark the model generation and checking for use in taint analysis. {\bf Properties} indicates the number of taint properties (e.g., user input flowing to logs) being analyzed.
  \textbf{Sources} indicates the number of sources of tainted data in the code and \textbf{Flows} indicates the number of flows where tainted data flows to a sink.}\label{table:benchmarks}
  \vspace{-10pt}
  \centering
  \rowcolors{2}{gray!15}{white}
  \begin{tabular}{clccc}
    \toprule
    Repository & Description & Properties & Sources & Flows \\
    \midrule
    sample           & CLI for encryption and file storage & 1 & 6 & 6 \\
    amazon-ssm-agent & Agent to manage AWS EC2 instances & 2 & 419 & 1 \\
    badger           & Key-value database in Go & 1 & 36 & 28 \\
    govatar          & Avatar generation library & 5 & 105 & 18 \\
    ofxgo            & Library for parsing and querying OFX & 2 & 125 & 5 \\
    prometheus       & Monitoring system for the cloud & 5 & 93 & 2 \\
    \bottomrule
  \end{tabular}
\end{table}

To answer these questions, we collect a set of applications and libraries written in Go. For each of those, listed in \Cref{table:benchmarks}, we define taint flow properties depending on the application's purpose. For example, a CLI tool should check that data that is direct user input is never used for executing code without being sanitized first. Naturally, since we are not the authors of the code we analyze, we may define problems that are unrealistic; in particular, the flows reported are not indicative of actual security issues. However, they can still serve as a means to evaluate the precision of our model generation. When designing the taint analysis scenarios, we manually verify the results to distinguish between false positives and real taint flows. We only keep the scenarios where the analysis returns zero to a few actual flows, where we observe no false positives, and where the analysis performs deep call-graph explorations (i.e., the data flows are non-trivial).
This ensures that when we perform the same taint analysis using the generated models, we can immediately flag new false positives (or false negatives). Since each modeled method participates in taint flow, we can quantify the variation in precision compared to the baseline.

We instantiate the baseline taint analysis with the \textsc{Argot}~\cite{argot} static analysis tool. The baseline taint analysis requires a pointer analysis to resolve aliases, which \textsc{Argot} includes as well. In particular, the baseline taint analysis implemented in \textsc{Argot} first performs a whole-program context-insensitive pointer analysis before running the taint analysis. The instantiation of our approach's soundness checking algorithm uses the same pointer analysis and taint analysis as the baseline; i.e., it is built on top of \textsc{Argot}.

In total, we collect \numTopLevelModels{} interesting methods to model. Recall that we support modeling interface methods, which means that checking a model sound may involve analyzing many top-level methods, all the implementations of the interface. This results in checking the soundness of taint flow models for \numTopLevelFuncsAnalyzed{} total methods.

\paragraph{Evaluation-specific assumptions}
The Go standard library uses features that the analyses of our soundness checking algorithm cannot soundly reason about, such as reflection and unsafe memory manipulation. To avoid false-positive unsound results in our evaluation due to these features, we use \textsc{Argot}'s predefined taint flow models for standard library functions. We assume that these models are sound. However, the coverage of those models is not complete, and our analysis flags very explicitly usages of unsound features when it encounters them.

\paragraph{Experimental setup}
We run all of our experiments on a machine with 4 vCPUs running Fedora 36 and with 64 GB of RAM. \textsc{Argot}~\cite{argot} uses some multi-threading during initialization, but all our analyses are single-threaded. Our agent is implemented using the Strands Agents~\cite{strands} framework and uses Claude Sonnet 4.5 for the LLM.

For each repository, we start from a configuration file specifying the taint analysis properties to prove, and a list of interesting methods (collected as described above) whose taint flows need to be modeled. We give our LLM agent as input the list of interesting methods and instructions on how to load the program to be analyzed. The agent is then tasked with generating a sound and precise taint flow model for each interesting method listed. While the agent has access to the soundness checker, it may decide not to use it. Therefore, we run the checker as a separate step after the agent has generated a model for every interesting method in the list. Three steps remain to complete the evaluation: (i) running the constructive analysis to attempt to build another version of the models, (ii) running the taint analysis with the agent-generated models, and (iii) running the taint analysis without any generated models.

\paragraph{\cref{rq:llm-effectiveness}: Soundness of the generated models.}
\begin{wraptable}{r}{0.6\textwidth}
  \caption{Model soundness checking success rate for the taint models generated by our agent. Each row lists the total number of top-level methods checked against the agent-generated models, and the ratio of sound/soundy/unsound outcomes for those checks.
  }
  \label{table:soundness-success}
  \centering
  \rowcolors{2}{gray!15}{white}
  \centering
  \small
  \begin{tabular}{l c c c c}
    \toprule
    Repository & Total & Sound & ``Soundy'' & Unsound \\
    \midrule
    sample & 551 & 99.9\% & 0.1\% & 0\% \\
    amazon-ssm-agent & 1952 & 91\% & 2\% & 6\% \\
    badger & 11 & 36\% & 64\% & 0\% \\
    govatar & 6 & 100\% & 0\%  & 0\% \\
    ofxgo & 3 & 33\% & 66\% & 0\%  \\
    prometheus & 12 & 100\% & 0\% & 0\% \\
    \bottomrule
  \end{tabular}
\end{wraptable}

\Cref{table:soundness-success} presents the result of checking the soundness of the agent-generated models. The three possible outcomes are \emph{sound}, \emph{unsound} or \emph{soundy}.
The taint flow model is \emph{sound} when the checker can prove that all the must-not-flows do not occur, and the analyses used for this proof do not encounter any language feature that threatens the soundness of the result. When the analyses can prove all the must-not-flows, but the method uses a feature that threatens the soundness result (see \Cref{sec:instantiation}), we list the outcome as \emph{soundy}~\cite{livshitsDefenseSoundinessManifesto2015}. An \emph{unsound} outcome means that our approach was not able to prove all of the must-not-flows; in this case, we have a form of counterexample to the model in the must-not-flow that may exist, although it may be a false positive.

During our evaluation, we encountered only 118 methods (5\%) for which the agent generated an \emph{unsound} model. This may sound surprising given our agent has access to the soundness checking tool, and thus should always check its results. But the fact that this happened only for the repository with largest number of function to check hints at the cause; the agent gave up on making some of the models sound because of the size of the task. In this particular occurrence, it was for an interface method summary with a large number of implementations. A possible fix would be to have the agent only consider a few summaries at a time.

There was a smaller proportion of \emph{soundy} outcomes (2\%), although the ratio of soundy to sound is more concerning for some of the repositories (ofxgo and badger). In those two cases, we manually checked that the models were sound. 

Our approach could not show the summary was unsound, but found unsound language features 
(e.g., reflection and unsafe) in the call graph rooted at the method
 whose taint flow model was ``soundy''. We observed that the agent was particularly careful when the checking tool reported such unsound features and often justified its decision of still generating the ``soundy'' model despite the warnings.


\paragraph{\cref{rq:checker-ablation}: Usefulness of soundness checking sub-analyses}
Since we instantiate our soundness checking approach with specific static analyses, we evaluate whether each analysis is useful by recording the number of times our soundness checking approach was useful in deciding the soundness of a model;
\Cref{table:check-method-used} presents our results. Each analysis was used at least once. In particular, our approach's ability to deduce callee summaries and recursively check their soundness was crucial in checking the soundness of the LLM-generated models for the amazon-ssm-agent repository. The LLM agent ended up generating the most-general summary for over 2,000 methods across all the applications we evaluated. However, this does not imply that the analysis can be replaced by simply using most-general summaries. The agent generated those summaries only where appropriate; had it used most-general summaries too frequently, we would expect additional false positives in the taint analysis, which we did not observe in our evaluation of \cref{rq:llm-effectiveness}.
\begin{wraptable}{r}{.6\textwidth}
  \caption{
    Model soundness checking analyses used when checking the models generated by the agent. Each analysis (most \textbf{gen}eral,
    \textbf{typ}e, \textbf{immut}ability, \textbf{read}, and \textbf{rec}ursive) is counted once per method being analyzed when it contributes to the outcome of the check.
  }
  \label{table:check-method-used}
  \centering
  \rowcolors{2}{gray!15}{white}
  \centering
  \small
  \begin{tabular}{l ccccc}
    \toprule
    Repository & {\bf gen.} & {\bf typ.} & {\bf immut.} & {\bf read} & {\bf rec.} \\
    \midrule
    sample & 547 & 0 & 5 & 0 & 0 \\
    amazon-ssm-agent & 1800 & 12 & 48 & 0 & 118 \\
    badger & 6 & 2 & 4 & 0 & 0 \\
    govatar & 5 & 1 & 2 & 0 & 0 \\
    ofxgo  & 1 & 0 & 2 & 1 & 0 \\
    prometheus & 4 & 5 & 7 & 0 & 0 \\
    \bottomrule
  \end{tabular}
\end{wraptable}

\paragraph{\cref{rq:llm-effectiveness}: Precision of Model Generation}
When generating a taint flow model, an LLM could simply choose the most-general summary. Our approach prompts the LLM agent to generate a model of sufficient precision for a generic taint analysis to succeed without too many false positives. Crucially, our approach does not ensure the \emph{most precise} taint flow models; we evaluate whether the LLM-generated taint flow models are \emph{precise enough} to not result in false positives in real-world taint analysis scenarios.

We observe that for all our benchmarks, no additional false positives are returned by the taint analyzer when using the LLM-generated taint flow models compared to without them. In order to obtain the baseline of false positives for the taint analysis examples where the analysis times out without summaries, we ran the analysis with carefully hand-written models for some of the methods that take a significant amount of time to analyze.

\todo{Add an evaluation which substitutes each LLM-generated model with the most-general summary to make sure that the most-general summary results in more false-positives.}

\begin{table}
  \caption{Running time of taint analysis without any models (baseline), taint analysis with LLM-generated sound/soundy models, model generation using the LLM agent, soundness checking step, and sound model generation via the constructive approach. All times are in seconds. TO indicates time-out after 30 min. \todo{Define $\dagger$}}
  \label{table:runtime-comparison}
  \centering
  \small
  \rowcolors{2}{gray!15}{white}
  \begin{tabular}{l c c | c c c}
    \toprule
    Repository & Baseline & With Summaries & Summarization & Check & Constructive \\
    \midrule
    sample & 26.03 & 26.02 & 521.85 & 9.93 & TO \\
    amazon-ssm-agent & TO & 130.35 & 424.01 & 43.17 & TO \\
    badger & 47.04 & 26.27 & 355.95 & 15.7 & 57.06$\dagger$ \\
    govatar & 57.05 & 54.29 & 253.15 & 2.17 & 2.75$\dagger$ \\
    ofxgo & TO & 9.27 & 244.16 & 7.23 & 66.54$\dagger$ \\
    prometheus & TO & 69.81 & 743.78 & 141.02 & 184.14$\dagger$ \\
    \bottomrule
  \end{tabular}
\end{table}

\paragraph{\cref{rq:workflow-efficiency}: Efficiency of Checking and Taint Analysis}
For a fair comparison, \Cref{table:runtime-comparison} reports the running time of all the components of our evaluation. Generating the models with an LLM agent should likely run only once, but a user should check the soundness of those models before each run of the taint analysis, as our approach can only check that the models are sound with respect to a given program.
The efficiency questions are then: (i) is running the soundness checker and the taint analyzer using the LLM-generated taint flow models faster than running the taint analyzer alone without any models? (ii) is running the LLM-based summarization and the soundness checker faster than constructing the models using the taint analysis?

To answer (i), we note that the baseline taint analysis times out after 30 minutes for 3 out of 6 applications; yet with the LLM-generated models that we prove sound, the slowest taint analysis run with the soundness check takes less than 3 minutes.
The only case where the combination of the soundness check and taint analysis with summaries takes longer than the baseline is for the sample repository, where both take less than 1 min.
Running the soundness checking step alone is efficient: the slowest is for prometheus, which took slightly over 2.5 minutes. In conclusion, it is reasonable to check the soundness of the LLM-generated models alongside the taint analysis or each time the program changes.

To answer (ii), we note that the constructive approach times out for both the sample and amazon-ssm-agent programs because they contain methods with complex intra-procedural taint flows, causing a state space explosion. For the remaining programs, the methods we model have simpler taint flows, so the constructive approach completes quickly; however, it still only produces sound summaries for less than 10\% of the methods. We allow partial generation by timing out on individual summaries.

Our soundness checker requires solving a Max-SAT problem to deduce callee summaries. However, even with field-sensitive taint flow models, we did not observe any scalability issues for this component of our approach because intra-procedural flows have a limited size. This may become an issue if we evaluate methods with hundreds of inputs and outputs, but such methods are rare in real-world codebases. For example, checking the soundness of a taint flow model for the \code{client.NewRequest} method in the amazon-ssm-agent codebase resulted in 956 total Max-SAT constraints, which the solver solved in 165 microseconds, resulting in a deduced callee model with 297 field-sensitive flows. Therefore, while our approach is theoretically NP-Complete in the worst case, we observed no scalability issues in the soundness checker in our evaluation.





%% file: sections/7_related_work.tex
\section{Related Work}


\paragraph{Compositional analysis}

The foundational work by Sharir and Pnueli~\cite{sharirTwoApproachesInterprocedural1978} proposes the functional approach to data flow analysis, enabling a composable context-sensitive inter-procedural analysis by summarizing the effects of methods. Summary-based functional approaches such as IFDS~\cite{repsPreciseInterproceduralDataflow1995}, IDE~\cite{sagivPreciseInterproceduralDataflow1995}, and WPDS~\cite{repsWeightedPushdownSystems2005} enable compositional inter-procedural dataflow analysis.

\paragraph{User-defined taint flow models}

Some analysis tools support user-defined taint flow models for methods that cannot be efficiently analyzed. Java taint analyses such as TAJ~\cite{trippTAJEffectiveTaint2009}, \textsc{Andromeda}~\cite{trippAndromedaAccurateScalable2013}, FlowDroid~\cite{arztFlowDroidPreciseContext2014}, and \textsc{CompTaint}~\cite{banerjeeCompositionalTaintAnalysis2023}, and the Go taint analysis \textsc{Argot}~\cite{argot}, use user-provided models for complex library methods. However, these approaches place the burden of correctness on the user---they do not verify that the provided models are sound. Our work addresses this limitation by providing a technique to verify the soundness of these user-provided models.

\paragraph{Synthesizing taint flow models}

Approaches that synthesize taint flow models without performing a static taint analysis do not check or enforce the soundness of the models.
\textsc{Taser}~\cite{staicuExtractingTaintSpecifications2020} uses a dynamic taint analysis to extract taint flow models of JavaScript methods, but dynamic taint analysis is unsound.
Zhai et al.~\cite{zhaiC2STranslatingNatural2020} use syntax guided synthesis to extract taint summaries from Java library method documentation but they do not analyze source code and have no soundness guarantees.
\textsc{STaint}~\cite{jiSTaintDetectingSecondOrder2025} prompts an LLM to infer taint flow models of database operations in PHP applications based on the function's source code, the taint analysis state at the call site, and the schema.
Our approach uses an LLM \emph{agent} equipped with tools to increase the precision of the taint flow models it synthesizes. We also check the soundness of the synthesized models.

Some approaches compute sound models with respect to a known program but have significant limitations.
\textsc{StubDroid}~\cite{arztStubDroidAutomaticInference2016} (Java) and \textsc{ModAlyzer}~\cite{schubertLosslessPersistedSummarization2021} (C/C\texttt{++}) compute complete and reusable taint models via an inter-procedural taint analysis and support higher-order methods, but also inherit the scalability limitations of taint analyses. 

Approaches that infer taint models via abduction~\cite{zhuAutomatedInferenceLibrary2013} or solving a context-free language reachability problem~\cite{bastaniSpecificationInferenceUsing2015} are similar to our callee summary deduction approach by synthesizing sound summaries with respect to known flows. However, these approaches assume that the method implementations are unavailable, thus needing a human oracle to check soundness, whereas our approach checks the soundness of the deduced callee summaries.
\textsc{Flowistry}~\cite{crichtonModularInformationFlow2022} soundly infers explicit and implicit \emph{information flow} (which includes implicit taint flows) within Rust functions using their type signatures. The Rust type system encodes ownership and mutability restrictions for references, which is necessary for \textsc{Flowistry}. Our work applies to languages other than Rust by verifying arbitrary taint flow models using type \emph{and} may-alias information which requires access to the program's source code. However, our approach is limited to only explicit taint flows.

\paragraph{Light-weight static analyses}

There is an established body of work exploring the relationship between mutability and data flow, starting from the program dependence graph~\cite{ferranteProgramDependenceGraph1987}.
FlowDroid~\cite{fritzHighlyPreciseTaint2013} avoids tainting callee parameters that have an immutable type in Java (e.g., \code{String}) but does not perform an immutability analysis to handle parameters that have a mutable type but are never modified in the callee.
\textsc{Flowistry}~\cite{crichtonModularInformationFlow2022} uses Rust's type system, which encodes the mutability of a function's outputs in its type signature, to increase the precision of its flow analysis.

FlowCFL~\cite{milanovaFlowCFLGeneralizedTypebased2020a} combines context-free language (CFL) reachability with \emph{reference immutability} to increase the precision of inter-procedural data flow analyses. Our immutability analysis uses similar ideas but we do not perform CFL reachability to construct taint flow summaries as they are calling-context-insensitive.
Our immutability analysis is most similar to the side-effect analysis of Milanova et al.~\cite{milanovaParameterizedObjectSensitivity2002}.


%% file: sections/8_conclusions.tex
\section{Conclusion}

We present an approach that allows taint analysis users to analyze their applications more efficiently with the help of LLMs, without compromising soundness or precision.
While we use an LLM agent to synthesize precise taint flow models, it would be interesting to investigate more efficient symbolic or machine learning techniques, or apply our soundness checking algorithm to verify existing approaches~\cite{staicuExtractingTaintSpecifications2020, zhaiC2STranslatingNatural2020, jiSTaintDetectingSecondOrder2025}.
Our soundness checking algorithm may be extended with other analyses, and, in parallel, our existing analyses may be extended to better make use of the precision requirement expressed through the generated models.
We only instantiate our approach for Go and future work could apply it to other languages such as Java.
We hope that our work leads to more research on using specification synthesis to scale static analysis while maintaining a high bar for soundness.


%% file: appendix/llm.tex
\section{Dataflow Model Synthesis Agent}\label{app:agent}

\subsection{Prompt}

\emph{We omit details of the prompt that are only relevant to the instantiation of our approach.}

\paragraph{The prompt.}

You are tasked with generating dataflow summaries for Go functions to be used with the Argot taint analysis tool. A dataflow summary describes how data flows through a function - which inputs affect which outputs.

\subsubsection{Complex Data Flows}

For flows through struct fields, arrays, or maps, use parentheses and suffixes:

\begin{itemize}
    \item \code{(!arg 0).field} - Flow through struct field
    \item \code{(!ret 0)[*]} - Flow through array/slice/map elements
    \item \code{(!arg <x>).field1[*].field2} - Nested field and element access
    \item For maps, \code{[*]} represents flow to both keys and values (no distinction possible).
\end{itemize}

\subsubsection{Task Instructions}

\begin{enumerate}
    \item **Analyze the function**: Understand what the function does and how data moves through it
    \item **Identify flows**: Determine which inputs affect which outputs
    \item **Be conservative**: Include all possible data flows to ensure soundness
    \item **Consider all paths**: Account for all execution paths (branches, loops, error cases)
    \item **Handle mutations**: If arguments are modified, include flows back to them
\end{enumerate}

\subsubsection{Important Notes}

\begin{itemize}
    \item **Soundness**: Your summary must be conservative - include all possible flows
    \item **Validation**: You can use \texttt{argot\_dataflow\_check} for summaries written in configuration files
    \item **Error handling**: Consider flows through error return values in Go's \code{(result, error)} pattern
    \item **Mutations**: If the function modifies input arguments, include flows back to them
    \item **All execution paths**: Consider branches, loops, and error conditions
\end{itemize}

\subsubsection{Example Request}

``Generate a dataflow summary for function \code{thisFunctionIsCalled} in file \texttt{path/to/sample.go} in package \code{main}.''

``Generate dataflow summaries for all the function suggested by the taint tool running with the configuration \texttt{argot-config.yaml}''

``Generate only the YAML dataflow summary for the provided function. Do not include explanations unless the flow is complex and needs clarification.''